\documentclass[showpacs,aps,prx,twocolumn,superscriptaddress,longbibliography]{revtex4-2}

\usepackage[english]{babel}
\usepackage[T1]{fontenc}
\usepackage[utf8]{inputenc}
\usepackage{amsmath}
\usepackage{amsthm}
\usepackage{amsfonts}
\usepackage{dsfont}
\usepackage[normalem]{ulem}
\usepackage{xr-hyper}
\usepackage{hyperref}
\usepackage{amssymb,graphicx}
\usepackage{tikz}
\usetikzlibrary{quantikz2}
\usetikzlibrary{decorations.pathmorphing}
\usepackage{circuitikz}
\usepackage{blochsphere}
\usepackage{siunitx}
\usepackage{svg}
\usepackage{mathbbol}
\usepackage{amsmath}    
\usepackage{braket}
\usepackage{verbatim}
\usepackage{tabularx}
\usepackage{multirow, array, booktabs}
\usepackage{algorithm2e}
\usepackage{algpseudocode}
\RestyleAlgo{ruled}
\usepackage{tabularx}
\usepackage{subfig}% subfigure package
\usepackage{booktabs}
\usepackage{multirow}
\usepackage{graphicx}
\newsavebox\MBox

\newcommand*{\colorboxed}{}
\def\colorboxed#1#{%
  \colorboxedAux{#1}%
}
\newcommand*{\colorboxedAux}[3]{%
  \begingroup
    \colorlet{cb@saved}{.}%
    \color#1{#2}%
    \boxed{%
      \color{cb@saved}%
      #3%
    }%
  \endgroup
}

\begin{document}
%
%%%%%%%%%%%%%%%%%%%%%%%%%%%%%%%%%%%%%%%%%%%%%%%%%%%%%%%%%%%%%%%%%%
%Title of the paper
%%%%%%%%%%%%%%%%%%%%%%%%%%%%%%%%%%%%%%%%%%%%%%%%%%%%%%%%%%%%%%%%%%
%
%\title{Noisy Quantum Walk (search) on the cycle (and torus) via circuit Quantum Electrodynamics QCA simulations}
\title{Noisy simulations of Quantum Walk and Quantum Walk search via Quantum Cellular Automata on a semiconducting spin processor emulator}
%
%%%%%%%%%%%%%%%%%%%%%%%%%%%%%%%%%%%%%%%%%%%%%%%%%%%%%%%%%%%%%%%%%%
%List of Authors
%%%%%%%%%%%%%%%%%%%%%%%%%%%%%%%%%%%%%%%%%%%%%%%%%%%%%%%%%%%%%%%%%%
%
\author{A.~Mammola}
\affiliation{C12 Quantum Electronics, Paris, France}
%\affiliation{Aix-Marseille Universit\'{e}, LIS, Marseille}
\affiliation{Aix Marseille Univ, CNRS, LIS, Marseille, France}
\author{Q.~Schaeverbeke }
\affiliation{C12 Quantum Electronics, Paris, France}
\author{G.~Di Molfetta}
\affiliation{Aix Marseille Univ, CNRS, LIS, Marseille, France}
\date{\today}
%%%%%%%%%%%%%%%%%%%%%%%%%%%%%%%%%%%%%%%%%%%%%%%%%%%%%%%%%%%%%%%%%%
%Paper abstract
%%%%%%%%%%%%%%%%%%%%%%%%%%%%%%%%%%%%%%%%%%%%%%%%%%%%%%%%%%%%%%%%%%
%
\begin{abstract} 
%
%MISSING MOTIVATION
%
In this work we map NISQ-friendly implementations of the non-interacting QCA to a circuit Quantum Electrodynamics (cQED) hardware. 
We perform both noiseless and noisy simulations of the QCA one particle sector, namely the Quantum Walk, on $N$-cycles and $N \times N$ torus graphs. Moreover, within this framework, we also investigate the search problem and present a circuit for preparing the W state (i.e., the Dicke state with hamming weight one) using only N-1 $\sqrt{\text{iSWAP}}$ gates and no ancilla qubits. The noiseless simulations are conducted with the Qiskit Aer simulator, while the noisy simulations with C12 Quantum Electronics' in-house noisy emulator, \textit{Callisto}. 
We benchmark the performance of our implementations by analyzing the simulations via relevant metrics and quantities such as the state count distributions, the Hellinger Fidelity,  the $\ell^{1}$ distance, the hitting time, and success probability. Our results demonstrate that the QCA framework, in combination with cQED processors, holds promise as an effective platform for early NISQ implementations of Quantum Walk and Quantum Walk Search algorithms.
\end{abstract}

\maketitle

\section{Introduction}\label{sec:intro}
Proving quantum advantage within the Noisy Intermediate Scale Quantum (NISQ) \cite{Preskill} era is among the main challenges currently faced by the quantum computing community. Indeed, crafting a successful NISQ implementation is hard and requires suitable applications and algorithms, NISQ-driven algorithmic formulations, and hardware tailored implementations. 
Among the category of suitable algorithms, one can certainly find those rephrased in terms of the Quantum Walk (QW) \cite{QW}, the quantum equivalent of the classical random walk, which has been theoretically proven to bring a speedup, depending on the nature of the graph, e.g. exponential speedup on glued trees \cite{Childs_exp_speedup}. %with whom it shares the fact of finding various applications among different fields. 
Within the quantum formalism, random walks can be formulated within a Discrete or Continuous-time framework and exhibit simple properties, 
featuring a predictable phenomenology, a clear scaling with the system size and showcasing explicit quantum advantage \cite{Georg1}. 
Moreover, much like their classical counterparts, Quantum Walks (QWs) have diverse applications across fields such as quantum computing, quantum simulation, quantum information, and graph theory \cite{QW_review}. In the latter category, the Quantum Walk Search \cite{Shenvi} stands out as a reference application of QWs to graph-based search problems, requiring only a slight modification to the standard QW implementation.

The aforementioned properties, have rendered discrete and continuous-time QW and QW search algorithms popular benchmarking tools, respectively for digital and analog NISQ implementations, with many examples already present in literature \cite{Georg1, Georg2, Cycle, Razzoli, Portugal}.
In particular, Discrete-time QW (DTQW) \cite{Aharonov} implementations provide valuable insight into the quality of the many gate-based NISQ processors and emulators, which are currently being developed within the industry. Even so, such digital implementations are typically translated into circuits with considerable size and depth, %for two-qubit gates, which are very noisy and represent the main challenge to overcome for NISQ devices. 
thus restricting the number of time steps that can be accurately executed on NISQ devices.

Moreover, QWs are known to be the one-particle sector of Quantum Cellular Automata (QCA) \cite{Watrous}.
QCA are lattice-based universal computational models that evolve in discrete time steps under a unitary, local, and translation-invariant operator \cite{Arrighi, Notarnicola, QCA_Cirac},
making them \textit{close-to-physics} models.
As a result, QCA-based circuit implementations often facilitate the translation of quantum algorithms into straightforward experimental realizations, typically using low-connectivity lattices and two-qubit gates only. Comprehensibly, these features are particularly valuable for early NISQ simulations, as well as for applications in searching and optimization problems, distributed quantum computing \cite{Arrighi}, quantum field theory simulations \cite{Schwinger, Rydberg} and quantum error correction \cite{QCA_QEC}.
Different QCA designs can give rise to different DTQW dynamics and among these we can find schemes where the QCA evolution operator is only built with the two-qubit XY-gate \cite{Costa}. 
From an experimental point of view, the XY-gate is derived from the XY interaction \cite{XY}, which emerges, among many, in Josephson charge qubits linked by Josephson junctions \cite{Joseph}, nuclear spins interacting through a 2D electron gas \cite{Joseph2} and quantum dot spins coupled by a cavity \cite{quantum_dot_QED}. In particular, hardware based on the latter, such as the \textit{circuit Quantum Electrodynamics (cQED)} processor of C12 Quantum Electronics \cite{C12}, have recently attracted significant attention due to their potential to enable highly connected architectures—a rare and valuable feature among NISQ devices.
With their high connectivity and the integration of the XY-gate into their native gate set, cQED processors like C12's are ideal candidates for testing and being benchmarked by QCA implementations.

In this work, we map QCA implementations of the QW and QW search over the cycle and torus to C12 Quantum Electronics cQED hardware. The QCA implementations are then executed with Qiskit software development kit (SDK) \cite{Qiskit}, both with Qiskit Aer noiseless simulator and \textit{Callisto} \cite{Callisto}, C12 Quantum Electronics \textit{in-house} noisy emulator.
The simulations act as a proof of concept, showcasing the feasibility of efficiently implementing Quantum Walks (QWs) and Quantum Walk search via QCA on cQED NISQ processors. Moreover, via the Hellinger Fidelity \cite{Hellinger} and suitable QW search quantities \cite{Zhang_search}, we study the impact of noise and benchmark the expected performance of C12’s processor and our QCA implementations with respect to other QW implementations over quantum machines or emulators \cite{Georg1, Acasiete, Cycle}. 
To the author's knowledge, this study provides the first evaluation of QW and QW search implementations via QCA on a NISQ emulator. Clearly, further investigation is required to test such implementations over a real cQED processor, in higher dimensions or over different graphs \cite{Childs_forget}, as algorithm subroutines \cite{QAOA_QW}, for state preparation \cite{Chang:2023nls} or as a starting point for more complex simulations of quantum physical theories, e.g. the Schwinger model \cite{Schwinger}, or quantum phases of matter \cite{Rydberg}.

This document is organized as follows. Section \ref{section:background} briefly introduces the non-interacting one-particle-sector QCA model simulating the discrete-time QW and QW search.
Section \ref{section:cqed} provides a mapping of the QCA models introduced to a cQED hardware, with the specific example of C12 Quantum Electronics processor. 
Section \ref{section:Sims} provides a description of the algorithm and the software used to perform the QW and QW search simulations over cycles and torus graphs, with a discussion of the noisy simulation results performed with C12's \textit{in-house} emulator \textit{Callisto}. 
Finally, in \ref{section:concl} we summarize the results obtained in this work and provide an outlook for future directions to be explored.
\section{Model}\label{section:background}
A quantum walker moves on a d-dimensional graph under the action of a coin operator, acting on the walker's internal state encoding for its direction, and a coin-conditioned shift operator, acting on the lattice degrees of freedom. The QW dynamics is known to be simulated by the non-interacting one-particle sector QCA.  
Let us now describe a QCA model implementing the QW and QW search over one- and two-dimensional lattices with periodic boundary conditions, respectively cycles and torus graphs.

\begin{figure*}
    \centering
    \includegraphics[width=18 cm]{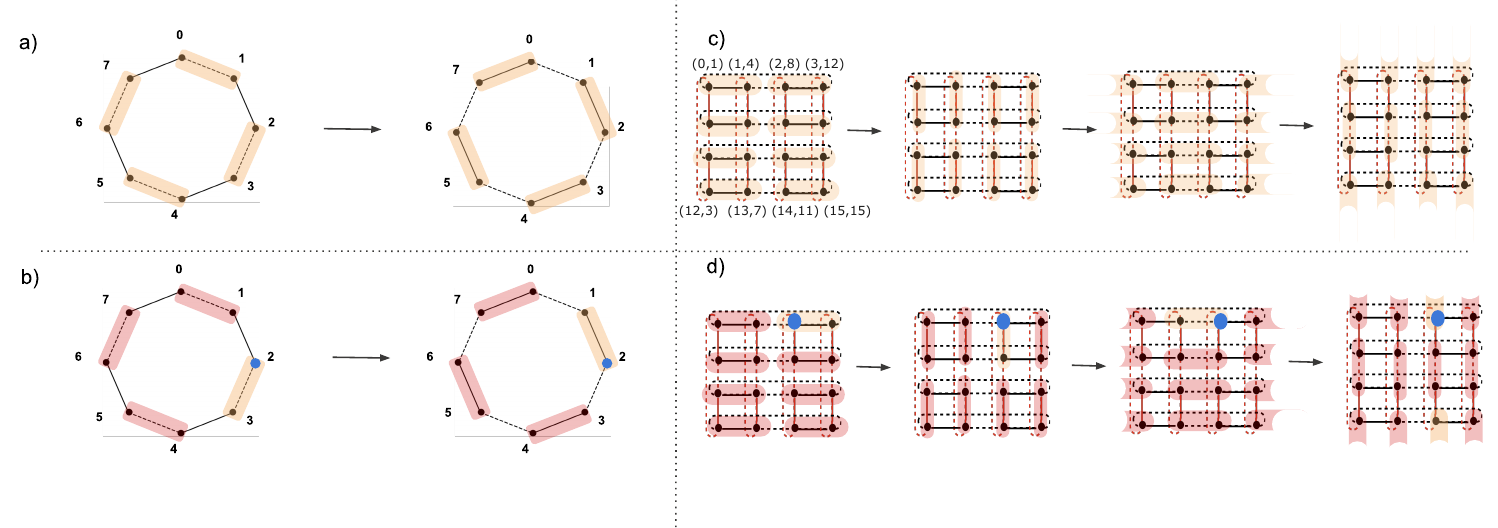}
    %\includesvg[width=19 cm]{im/cycle_torus_3_3}
    \caption{Left: One time-step of an 8-cycle QW (a) and QW search (b). The minimal tessellation cover of cycles is two: $\mathcal{T}_{0}$ (dashed line) and $\mathcal{T}_{1}$ (solid line), see Eq. (\ref{eq:tess_1D}). Each time-step is divided into two sub-steps, where two-qubit matrices $W_{j, j+1}(\theta_{j}=\frac{\pi}{4})$ (orange box) and $W_{j, j+1}(\theta_{j}=\frac{\pi}{2})$ (red box)  are applied via the transition functions $W_{\mathcal{T}_{0}}$ and $W_{\mathcal{T}_{1}}$, associated to each tessellation, see Eq. (\ref{eq:QCA_tr_func}).
    Right: One time-step of a 4x4 torus QW (c) and QW search (d). The minimal tessellation cover of torus graphs is four: $\mathcal{T}_{0_{x}0_{y}}$ (black solid line), $\mathcal{T}_{1_{x}0_{y}}$ (black dashed line),  $\mathcal{T}_{0_{x}1_{y}}$ (red solid line) and  $\mathcal{T}_{1_{x}1_{y}}$ (red dashed line), see Eq. \ref{eq:tess_2D}. For clarity, only the indices of the first and last rows of the first torus in panel (c) are shown. A single time-step is divided into four sub-steps, where two-qubit matrices $W_{j, j+1}(\theta_{j}=\frac{\pi}{4})$ (orange box) and $W_{j, j+1}(\theta_{j}=\frac{\pi}{2})$ (red box) are applied via the transition functions $W_{\mathcal{T}_{0_{x}0_{y}}}$, $W_{\mathcal{T}_{1_{x}0_{y}}}$, $W_{\mathcal{T}_{0_{x}1_{y}}}$ and $W_{\mathcal{T}_{1_{x}1_{y}}}$, associated to each tessellation see Eq. (\ref{eq:QCA_tr_func}). The blue dot in b) and d) indicates the marked vertex in the search.}\label{fig:cycles_torus}
\end{figure*}

Consider a N-cycle $\mathcal{C}(V_{\mathcal{C}}, \ E_{\mathcal{C}})$ and a $N\times N$ torus $\Gamma(V_{\Gamma}, \ E_{\Gamma})$ with fixed vertex sets $V\_$ and edge sets $E\_$, respectively associated to the Hilbert spaces $\{\mathcal{H}_{V_{\mathcal{C}}}, \  \mathcal{H}_{E_{\mathcal{C}}}\}$ and $\{\mathcal{H}_{V_{\Gamma}}, \  \mathcal{H}_{E_{\Gamma}}\}$. 
We label each vertex of the cycle $\mathcal{C}$ with $j \in \mathbb{N}$ and place there a qubit with the state $\ket{1}_{j}(\ket{0}_{j})\in \mathcal{H}_{V_{\mathcal{C}}}$, encoding for an (un)occupied site. In the case of the torus $\Gamma$, each vertex is labeled with (j,l), where $j \in \mathbb{N}$ on the x-axis and $l \in \mathbb{N}$ on the y-axis, and can be associated to the state $\ket{1}_{jl}(\ket{0}_{jl})\in \mathcal{H}_{V_{\Gamma}}$, encoding for an (un)occupied site. Within the torus, sites are sequentially labeled along each axis. 
For instance, in a $4 \times4$ torus, the first site of the first row is (0,0), the first site of the second row is (1,4) and the last site of the last row is (15,15). To simulate the QW dynamics, the cycle (torus) QCA has to be restricted to its one-particle sector, namely, for all sites j(l), only states of the likes $\ket{0_{0(0)}\cdots, 1_{j(l)}, \cdots 0_{N(N)}}$ are accepted. 
The graphs $\mathcal{C}(V_{\mathcal{C}}, \ E_{\mathcal{C}})$ and $\Gamma(V_{\Gamma}, \ E_{\Gamma})$ are divided into tessellations $\{\mathcal{T}_{k}\}_{k=0}^{K-1}$ \cite{Costa}. 
The global time-evolution operator of the QCA is then obtained as the product of unitary operators defined over each tessellation $\mathcal{W}=\prod_{k=0}^{K-1}W_{\mathcal{T}_{k}}(\theta_{k})$, 
where $W_{\mathcal{T}_{k}}(\theta_{k})= \bigoplus_{\alpha \in \mathcal{T}_{k}} e^{-i\theta_{k}H_{\alpha}}$ and $\alpha$ is a complete subgraph of G.  
As depicted in Fig. \ref{fig:cycles_torus}, the one and two-dimensional lattice minimal tessellation covers are respectively composed by the following two tessellations $\mathcal{T}_{k}$ 
\begin{equation}\label{eq:tess_1D}   
\begin{cases}
&\mathcal{T}_{0} = \{2j, \ 2j+1 \}\\
&\mathcal{T}_{1} = \{2j+1, \ 2j+2\} \\
\end{cases}
\end{equation}
and the following four tessellations $\mathcal{T}_{k_{x}k_{y}}$ 
\begin{equation} \label{eq:tess_2D}    
\begin{cases}
&\mathcal{T}_{0_{x}0_{y}} = \{(2j, l_{1}), \ (2j+1, l_{2})\}\\
&\mathcal{T}_{1_{x}0_{y}} = \{(2j+1, l_{1}), \ (2j+2, l_{2})\}\\
\\
&\mathcal{T}_{0_{x}1_{y}} = \{(j_{1}, 2l), \ (j_{2}, 2l+1)\}\\
&\mathcal{T}_{1_{x}1_{y}} = \{(j_{1}, 2l+1), \ (j_{2}, 2l+2)\}\\
\end{cases} \ ,
\end{equation}
where j and l $\in \mathbb{N}$. Each tessellation consists of the set of points $\{(j,l)\}$ satisfying a condition on the corresponding axis, while being independent with respect to the other axis. For example, an x-axis tessellation is determined by conditions on the x-coordinates while the y-coordinates are left trivial (denoted above by $l_{1}$ and $l_{2}$); the same holds for tessellations on the y-axis.
Accordingly, the cycle and torus QCA time evolution operators write
\begin{equation}\label{eq:QCA_tr_func}
\begin{split}
     \mathcal{W_{\mathrm{cycle}}}&=\mathcal{W}_{\mathcal{T}_{1}}\cdot \mathcal{W}_{\mathcal{T}_{0}}=\prod_{k=0}^{1}\bigg[\bigotimes_{j\in \mathcal{T}_{k}}W_{j, j+1}(\theta_{j})\bigg]\bigotimes_{j\not\in \mathcal{T}_{k}} \mathds{1}_{j}\\
     &=\bigg[\bigotimes_{j=0}^{N-1} W_{2j+1, 2j+2}(\theta_{j})\bigg]\cdot\bigg[\bigotimes_{j=0}^{N-1} W_{2j, 2j+1}(\theta_{j})\bigg]\\
\end{split}
\end{equation}
\begin{equation*}
    \begin{split}
\mathcal{W_{\mathrm{torus}}}&=\mathcal{W}_{\mathcal{T}_{1_{x}1_{y}}}\cdot \mathcal{W}_{\mathcal{T}_{1_{x}0_{y}}}\cdot \mathcal{W}_{\mathcal{T}_{0_{x}1_{y}}}\cdot \mathcal{W}_{\mathcal{T}_{0_{x}0_{y}}}\\
     &=\prod_{k_{x},k_{y}=0}^{1}\bigg[\bigotimes_{j\in \mathcal{T}_{k_{x}k_{y}}}W_{j, j+1}(\theta_{j})\bigg]\bigotimes_{j\not\in \mathcal{T}_{k_{x}k_{y}}} \mathds{1}_{j}\\
     &=\bigg[\bigotimes_{y=0}^{N-1} W_{2l+1,2l+2}(\theta_{y})\bigg]\cdot \bigg[\bigotimes_{x=0}^{N-1} W_{2j+1, 2j+2}(\theta_{x})\bigg]\\
     & \ \ \cdot\bigg[\bigotimes_{y=0}^{N-1} W_{2l,2l+1}(\theta_{y})\bigg]\cdot\bigg[\bigotimes_{x=0}^{N-1} W_{2j,2j+1}(\theta_{x})\bigg]\\
     \end{split}
\end{equation*}
where $\bigotimes_{j\not\in \mathcal{T}_{k}} \mathds{1}_{j}$ and $\bigotimes_{j\not\in \mathcal{T}_{k_{x}k_{y}}} \mathds{1}_{j}$ indicate that the transition function is acting as the identity on the sites that are not pertaining to the chosen tessellation (i.e. $j \not \in \mathcal{T}$). Note that these identities are omitted in the last term of the equations for notational simplicity, and that the subscript of the two-qubit matrix $W_{j, j+1}$ indicates only the qubit coordinates within the chosen tessellation (either along the x or y-axis). For example, $W_{2j,2j+1}$ with $x=0$ acts on the two qubits located at positions $(\mathbf{0}, 0)$ and $(\mathbf{1}, 4)$ respectively. The two-qubit gate
\begin{equation}\label{eq:QCA_QW_unit}
W_{j, j+1}(\theta_{j})=\begin{pmatrix}
                    1 & 0  & 0 & 0\\ 
                    0 & \cos{\theta_{j}}  & i\sin{\theta_{j}} & 0\\ 
                    0 & i\sin{\theta_{j}}  & \cos{\theta_{j}} & 0\\ 
                    0 & 0 & 0 & 1\\
    \end{pmatrix}\ ,
\end{equation}
is the XY-gate \cite{XY_2} and acts upon two neighboring qubits located in the x or y-axis of the lattice at position $(2j, \ 2j + 1)$ and $(2j+1, \ 2j + 2)$, as depicted in Fig. \ref{fig:cycles_torus}. 
The full unitary QCA evolution, generating the QW dynamics, for a single time-step writes $\ket{\psi(t+1)}=\mathcal{W_{\mathrm{cycle}}}\ket{\psi(t)}$ for the cycle and $\ket{\psi(t+1)}=\mathcal{W_{\mathrm{torus}}}\ket{\psi(t)}$ for the torus.
Various QWs can be implemented for different choices of $\theta_{i}$. We set $\theta_{i}=\frac{\pi}{4}$, for which the XY-gate is equal to $\sqrt{\text{iSWAP}}$ gate \cite{squared_root_iswap}.
For the QW search instead, given a marked vertex $v_{m}$, we set $\theta_{i}=\frac{\pi}{4}$, for $i = v_{m} $
and $\theta_{i}=\frac{\pi}{2} \ , \forall i \neq v_{m} $, for which the XY-gate equals the iSWAP gate.

As a natural consequence of QCA being \textit{physics-like} models, the QW implementations described above require $\mathcal{O}(n)$ qubits and $\mathcal{O}(nt)$  two-qubit gates (i.e. $\sqrt{\text{iSWAP}}$) with a depth of  $\mathcal{O}(t)$, where t is the time-step.
To date, several efficient circuit-based implementations of 1D QWs have been proposed. All methods require $\mathcal{O}(\log{N})$ qubits and differ in how the shift operator is implemented. Shakeel et al. \cite{Shakel} introduced a design using a Quantum Fourier Transform (QFT) based \textit{incrementer} and $\mathcal{O}(n^{2}t)$ C-NOT gates with a circuit depth of $\mathcal{O}(n t)$. Razzoli et al. \cite{Razzoli} later improved this by reducing the QFT usage, lowering the gate count and depth to $\mathcal{O}(n^{2})$ and $\mathcal{O}(nt+n^{2})$, respectively. 
Additionally, various efficient incrementers based on generalized C-NOTs have been proposed, including:

i) Gidney’s scheme \cite{Gidney} with Toffoli cost of $32n$ ($64n$ C-NOTs), depth of $\mathcal{O}(n)$ and one clean ancilla qubit ;

ii) Nie et al. \cite{nie2024quantumcircuitmultiqubittoffoli} with Toffoli cost $\mathcal{O}(n)$, depth $\mathcal{O}(\log^{2}{n})$ and one clean ancilla qubit ;

iii)Khattar and Gidney \cite{Gidney_2} with Toffoli cost $3n$, depth $\mathcal{O}(n)$ and $\log_{2}^{*}{n}$ clean ancilla qubits. 
\section{QCA mapping to cQED processor}\label{section:cqed}
As follows, we describe a mapping of the QCA described in the previous section to a Circuit Quantum Electrodynamics (cQED) processor, with the specific example of C12 Quantum Electronics hardware \cite{C12}. We start with a brief introduction of C12's processor, to then provide the QCA mapping.

cQED processors \cite{cottet2015, blais2021} are based on cavity Quantum Electrodynamics (QED) \cite{haroche1989}, which focuses electric fields in small regions around atoms to control light-matter interactions and generate entanglement. This concept can be applied to quantum circuits made from semi- or superconducting components, where artificial atoms (qubits) are embedded in a microwave cavity. The cavity acts as a quantum bus, enabling long-range entanglement by exchanging photons between qubits and thus, in principle, allowing for all-to-all connectivity. Within such setting, one-qubit gates are implemented with fast electric lines, while two-qubit gates are achieved by coupling qubits synchronously to the microwave cavity \cite{Dijkema_2024}.
\subsection{cQED semiconducting spin processor}
An example of a cQED processor is \textit{C12 Quantum Electronics} hardware, which stems from the single electron spin qubit paradigm \cite{burkard2023semiconductor}.
The particularities of C12's technology are twofold with respect to this paradigm.
The first one is the usage of a microwave resonator to mediate the coupling between the qubits in a cQED type architecture \cite{dijkema2025cavity}.
This allows to reach all-to-all connectivity in the processor at the cost of a new decoherence channel in the Purcell effect.
The second specificity is the usage of carbon nanotube as the host material for the spin qubit.
Carbon nanotubes, although less studied, have the advantage of being structurally simpler than the other materials typically used for spin qubits.
Indeed, while silicon or germanium will consist of layers deposited onto a substrate in which bidirectional confinement has to be performed, the quasi 1D structure of the tube has a natural confinement in one direction and therefore only requires confining the electron in the remaining direction.
On top of that, the tube can be suspended above the electrode used for the confinement, allowing for a molecular structure suspended in vacuum and consequently more isolated from its environment.

This type of cQED-based processor is promising but still requires further development to reach state-of-the-art capabilities.
Indeed, while the all-to-all connectivity offered by the resonator should, in principle, provide significant advantages for quantum computation, current chips do not achieve state-of-the-art coherence times \cite{yoneda2018quantum, mammola}.
Despite this, carbon nanotube–based spin qubits have already shown promise, currently holding the record for coherence times among cQED spin qubits \cite{neukelmance2025microsecond}.
At present, the primary challenge limiting the scalability of this platform in terms of two-qubit entanglement is achieving sufficient coupling with the resonator to enable high-fidelity gates. Progress on high-impedance circuit design is expected to enhance this coupling. Resonance between qubits constitutes another potential limitation; however, it can be actively controlled via electrostatic tuning of the double quantum dot and is therefore not anticipated to pose a major obstacle. Variations in decoherence rates across the qubit register may still affect gate fidelity depending on the specific configuration.
While promising, cQED semiconducting spin platforms still require further development before they can provide a quantum processor suitable for experiments.
Accordingly, the simulations presented in this work rely on theoretical projections using a hardware emulator with C12's proprietary noise model, which reflects the expected behavior of the processor under the assumption that the aforementioned technical limitations can be addressed in future experimental implementations.
Indeed, two-qubit gates on these platforms have only been demonstrated very recently, with fidelities around $80\%$ \cite{dijkema2025cavity}, highlighting the need for substantial optimization before practical experimental applications are feasible.
While precise timelines remain uncertain, it is reasonable to anticipate that quantum walk experiments with a few qubits could become feasible in the near future.
%
\begin{comment}
This type of cQED based processor is promising but is still requiring further development to reach state of the art capabilities.
%
Indeed, while on paper the all to all connectivity should offer a significant advantage for quantum computation, the chips containing a resonator do not achieve state of the art coherence times \cite{yoneda2018quantum, mammola}.
%
Despite that, carbon nanotube based spin qubits have already shown some promise as today they own the record for coherence time among cQED spin qubits \cite{neukelmance2025microsecond}.
%
While promising, cQED semiconducting spin platforms still require further development before they can provide a quantum processor suitable for running experiments. \textcolor{blue}{For this reason, the simulations presented in this work rely on theoretical projections using a hardware emulator with C12's proprietary noise model, assuming that the aforementioned technical limitations can be overcome in future experimental implementations.} In fact, two-qubit gates on these platforms have only been demonstrated very recently, with fidelities around $80\%$ \cite{dijkema2025cavity}, which still calls for significant optimization before experimental implementations can become practically relevant.
\end{comment}

C12's gate set includes the three Pauli rotations $R_{X}(\theta),\ R_{Y}(\theta), \ R_{Z}(\theta)$ with $\theta$ depending on experimentally controllable parameters and the gate-time, and the two-qubit XY-gate:
\begin{equation}\label{eq:C12_2qbt_gt}
 U^{(2)}(\theta=\gamma t_{g})=\begin{pmatrix}
                    1 & 0  & 0 & 0\\ 
                    0 & \cos{\theta}  & i\sin{\theta} & 0\\ 
                    0 & i\sin{\theta}  & \cos{\theta} & 0\\ 
                    0 & 0 & 0 & 1\\
    \end{pmatrix}\ ,
\end{equation}
where $t_{g}$ is the gate time and $\gamma$ the cavity mediated qubit-qubit coupling. For $t_{g}=\frac{\pi}{2 \gamma}$ the two-qubit matrix $U^{(2)}$ implements the iSWAP gate and for $t_{g}=\frac{\pi}{4\gamma}$ the  $\sqrt{\text{iSWAP}}$.  
\begin{figure}[htp]
  \centering
    \includegraphics[width=6 cm]{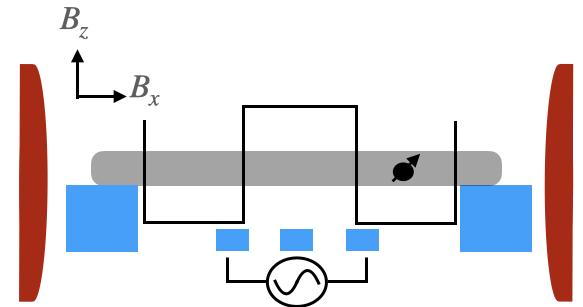}
 \caption{Schematic of C12's spin qubit. An electron is hosted in a carbon nanotube (grey tube),  suspended over five electrodes (blue squares) generating a double quantum dot (black squared well) trapping the particle (black arrowed dot), and immersed in an asymmetric magnetic field. One qubit gates are then realized by changing the dot's voltage (black wave) and two-qubit gates are realized by embedding two qubits within the same microwave cavity (red resonator plates).}
 \label{fig:c12_qubit}
\end{figure}
On top of this basis gate-set, to be operative, C12's multi-qubit processor needs ON-OFF gates, which decouple unused qubits from the resonator during an operation and sets them within the memory or $\textit{idle}$ state. Experimentally, this can be achieved by forcing the electron to be stuck in one of the dots, e.g. by applying a large bias voltage between the two dots, in a process known as the Landau-Zener transition \cite{Landau_Zener}. 
\subsubsection{Decoherence}
Decoherence affects the qubit's life time and the gate error rates, thus limiting the maximal depth of a quantum circuit executable on the processor.
To have a realistic description of the qubit and its operation, it is crucial to account for the sources of decoherence within the qubit’s external environment. We separate the sources into two categories based on their effect on the qubit, i.e. relaxation and dephasing, which respectively describe processes by which: i) the qubit relaxes towards its ground state over a time $T_{1}$ and ii) the qubits lose coherence without effect on the populations with time $T_{2}$. 
The main sources include the microwave cavity, modeled as a single-mode resonator, where qubits are influenced by the Purcell effect \cite{Purcell} leading to energy relaxation. This effect is determined by the cavity's quality factor, which depends on imperfections in the conductors used to build it, the electron-photon coupling, the qubit-cavity frequency shift and the electronics around the cavity. Another source are the elastic properties of Carbon Nanotubes, where phonon vibrational modes cause both dephasing and relaxation of the qubit state \cite{phonon1, phonon2, phonon3, phonon4,phonon5}. Charge noise also plays an important role. Defined as fluctuations in the qubit's electrostatic environment, it predominantly causes dephasing and also contributes to relaxation \cite{ch_noise1, ch_noise2, ch_noise3, ch_noise4}. Additionally, cross-talk errors arise from unwanted interactions during computation \cite{Sarovar2020}, such as residual idle-active qubit couplings emerging in cavity-mediated two-qubit gates, which could potentially restrict the hardware's connectivity.

Formally, we treat decoherence within the density matrix formalism \cite{Petruccione} and write the time evolution of the system via the Lindblad equation \cite{Lindblad:1975ef, Gorini}
\begin{equation}
    \dot{\rho} = -i [H_{g},\rho]+ K \mathcal{L}\bigg[\sum_{i}\sigma_{-}^{(i)}\bigg](\rho) + \delta \mathcal{L}\bigg[ \sum_{i}\sigma_{3}^{(i)}\bigg](\rho)
\end{equation}
where $\rho$ is the system's density matrix, $H_{g}$ the gate Hamiltonian, i.e. the Hamiltonian describing the system whose unitary is among C12's basis gate-set, $\mathcal{L}\bigg[\sum_{i}\sigma_{-}^{(i)}\bigg](\rho)$ and $\mathcal{L}\bigg[ \sum_{i}\sigma_{3}^{(i)}\bigg](\rho)$ are the Lindblad operators acting on the i-qubit, respectively encoding for the relaxation of the states with a rate $K$ and dephasing with a rate $\delta$.  The Lindblad operator is defined as $\mathcal{L}\big[A\big](\rho(t))=A\rho(t)A^{\dagger}-\frac{1}{2}AA^{\dagger}\rho(t)-\frac{1}{2}\rho(t) AA^{\dagger}$. 
\subsubsection{C12's emulator Callisto}
C12's \textit{in-house} emulator \textit{Callisto} is built on top of the Qiskit library with the noise model described in the previous section, allowing to perform realistic noisy simulations. %
The version of the emulator used for this work is $c12-callisto-clients==2.0.1$; it allows for all-to-all connectivity and has the basis gate-set and Fidelity reported in Table \ref{tab:1}.
\begin{table}[htp]
\centering
\begin{tabular}{|c|c|c|}
\hline
%\multicolumn{3}{|c|}{Basis gate set} \\
%\hline
Gate & Fidelity\\
\hline
$R_{X}(\theta), \ R_{Y}(\theta)$ & 0.9999 \\
%\hline$R_{Y}$ & 99.9989  & $ \approx 10^{-3}$ \\
\hline
$R_{Z}(\theta)$ & 1 \\
\hline
$U^{(2)}(\frac{\pi}{2})=\text{iSWAP}$ & 0.9987 \\%99.8731
\hline
$U^{(2)}(\frac{\pi}{4})=\sqrt{\text{iSWAP}}$ & 0.9991 \\%99.9165
%0.9939224266824571
\hline
\end{tabular}
\caption{\textit{Callisto} native gate-set and gate Fidelity}
\label{tab:1}
\end{table}

Leveraging the high connectivity of cQED hardware, the QCA described in section \ref{section:cqed} can be easily implemented with C12's qubits of Fig. \ref{fig:c12_qubit}.
The QW and QW search QCA dynamics is then reproduced by applying at each time step the native two-qubit gates $U^{(2)} (\theta=\gamma t_{g})$, respectively with gate times $t_{g}=\frac{\pi}{4\gamma}$ and $t_{g}=\frac{\pi}{2\gamma}$, as the unitaries $W_{2i, \ 2i+1}(\theta_{i}), \ W_{2i+1, \ 2i+2}(\theta_{i})$ of Eq.  \ref{eq:QCA_tr_func}. We note that the periodic boundary conditions impose that, at the borders of the one- and two-dimensional lattices, the two qubit matrix $W_{i, \ i+1}(\theta_{i})$ acts  on the last and first qubits of the row or column.
\section{Simulations}\label{section:Sims}
In the following, we will describe the algorithms and the software used to run noiseless and noisy simulations of the QW and QW search over the cycle and torus via a cQED QCA. In the second part, we will outline the metrics and quantities used to study and benchmark the QW simulations and then discuss the simulation results. 
\subsection{Algorithm} \label{sec:algos}
Algorithms (\ref{algo:algo1}, \ref{algo:algo2}) respectively describe the implementation of QW and QW search on cycle and torus graphs via QCA models mapped to cQED hardware. Following the mapping outlined in Section \ref{section:cqed}, a system of $N$ qubits is configured with the appropriate connectivity to represent either a cycle $\mathcal{C}(E_{\mathcal{C}}, V_{\mathcal{C}})$ or a torus $\Gamma(E_{\Gamma}, V_{\Gamma})$, enforcing periodic boundary conditions in both cases.
The qubits can be operated upon with the hardware's native gate-set of Table (\ref{tab:1}) and
the QW dynamics is generated with the time-evolution operator of Eq. (\ref{eq:QCA_tr_func}). The algorithms are divided into \textbf{initialization}, where the qubits are initialized in the required initial state, and \textbf{propagation}, where the QCA transition function is applied to the system $N_{stps}$ times and the final state $\ket{\psi_{N_{stps}}}$ is returned.

In the QW (Algorithm \ref{algo:algo1}), if \textit{'symmetric'} is \textbf{True} the system is initialized in a superposition of two chosen sites, whereas if \textit{'symmetric'} is \textbf{False} the system is simply initialized on one chosen site. Propagation is performed using the two-qubit gate $U^{(2)}(\frac{\pi}{4})$ from Eq. (\ref{eq:C12_2qbt_gt}).

In the QW search (Algorithm \ref{algo:algo2}), the \textit{Dicke\_one\_initializer()} function (Algorithm \ref{algo:algo3}), initializes N qubits in the Dicke $\ket{D^{N}_{k}}$ state with hamming weight $k=1$ , i.e. an equiprobable superposition of the one-particle sector states $\ket{\psi_{0}}=\sum_{i=0}^{N-1}\frac{1}{\sqrt{N}}\ket{\cdots, 0_{i-1}, 1_{i}, 0_{i+1}, \cdots}$, via repeated application of the two-qubit gate $U^{(2)}(\frac{\pi}{4})=\sqrt{\text{iSWAP}}$. Precisely, the circuit we present here requires $N-1$ $\sqrt{\text{iSWAP}}$ gates. A vertex $v_m$ is then marked: $\theta_{i}=\frac{\pi}{4}$ for $i=v_{m}$ and $i=\frac{\pi}{2} \ \forall i\neq v_{m}$, namely the transition function applies $U^{(2)}(\frac{\pi}{4})$ to unmarked sites and $U^{(2)}(\frac{\pi}{2})$ to the marked one.

An example QW and QW search circuit over an 8-cycle is shown in Fig. (\ref{fig:circuits}), appendix \ref{sec:app}.
\begin{algorithm}
    \KwData{
$N$ qubits with  all-to-all connectivity mapped to $N$-cycle $\mathcal{C}(E_{\mathcal{C}},V_{\mathcal{C}}) \ (N\times N$ torus $\Gamma(E_\Gamma,V_{\Gamma}))$ %, with corresponding tesselation cover, see Eq. (\ref{eq:tess_1D, eq:tess_2D}). 

%C12's gate-set (Table \ref{tab:0}), 
    }
    %\KwResult{Ideal number of qubit per cavity}
    
    \hrulefill
    
        \textbf{Initialization:}
    %\vspace{2 mm}
    
    \eIf{symmetric == True}{
    $\ket{\psi_{0}}=U_{i,i+1}(\frac{\pi}{4})^{(2)}\cdot R_{X_{i}}(\frac{\pi}{2})\ket{0_{0},\cdots, 0_{N-1}}$
    
    $\quad \quad =\frac{1}{\sqrt{2}}\ket{0_{0}, \cdots}(\ket{0_{i}}+\ket{1_{i}})\ket{\cdots, 0_{N-1}}$ \;}
    {$\ket{\psi_{0}}=R_{X_{i}}(\frac{\pi}{2})\ket{0_{0},\cdots, 0_{N-1}}$ \;
    
     %\quad \quad =\ket{0_{0}, \cdots, 1_{i},\cdots, 0_{N-1}}$ \;
    }
    
    \vspace{2 mm}
     $\theta_{i}=\frac{\pi}{4}$ $\forall i$ \;
    $\mathcal{W_{\mathrm{cycle}(\mathrm{torus})}}=\prod_{k(,l)=0}^{1}\bigg[\bigotimes_{i\in \mathcal{T}_{k(l)}}^{N-1}W_{i, i+1}(\theta_{i})\bigg]$ ,
    
    with $W_{i, i+1}(\theta_{i})=U^{(2)}(\frac{\pi}{4})$, Eq. (\ref{eq:C12_2qbt_gt}) \;
    
    \hrulefill
    
    \textbf{Propagation:}  
    
    \vspace{2 mm}
    
        \For{$(n=1,\ n<N_{stps})$:}{$\ket{\psi_{n}}=\mathcal{W}_{cycle(torus)} \ket{\psi_{n-1}}$ \;} 
    \vspace{2 mm}
        
        \textbf{Return} $\ket{\psi_{N_{stps}}}$ \;
    
    \caption{QW on N-cycles $\ (N \times N$ torus graphs) via cQED-QCA}\label{algo:algo1}
\end{algorithm}
\begin{algorithm}
    \KwData{ $N$ qubits with  all-to-all connectivity mapped to $N$-cycle $\mathcal{C}(E_{\mathcal{C}},V_{\mathcal{C}}) \ (N\times N$ torus $\Gamma(E_\Gamma,V_{\Gamma}))$.%, with corresponding tesselation cover, see Eq. (\ref{eq:tess_1D, eq:tess_2D}). 

%C12's gate-set (Table \ref{tab:0}), 
    }
    %\KwResult{Ideal number of qubit per cavity}
    
    \hrulefill
    
        \textbf{Initialization:}
    
                \vspace{3 mm}
                
                $\ket{\psi_{0}}=\textit{Dicke\_one\_initializer(N)}$, see Algorithm \ref{algo:algo3}\\
                $\ \ \ \ \ \ =\frac{1}{\sqrt{2^{N}}}\sum_{i=0}^{N-1}\ket{\cdots,0_{i-1},1_{i},0_{i+1},\cdots}$
                %Initializer(N, \ket{0_{0}, \cdots, 0_{N-1}})=$
                 \;

               \vspace{2 mm}
                mark vertex $v_{m}$ \ | 
         
             $\theta_{i}=\frac{\pi}{4}\ ,$ for $i = v_{m} $
             and $\theta_{i}=\frac{\pi}{2} \ , \forall i \neq v_{m}$ \;     
              
              \vspace{2 mm}
             
             $W_{i, i+1}(\theta_{i})=R_{Z_{i}}(-\frac{\pi}{2})\otimes R_{Z_{i+1}}(-\frac{\pi}{2})\cdot U^{(2)}(\frac{\pi}{2})$ 
             $\forall i \neq v_{m}$,

            $W_{i, i+1}(\theta_{i})=U^{(2)}(\frac{\pi}{4})$ for $i \ = v_{m}$, Eq. (\ref{eq:C12_2qbt_gt}) \;
               $\mathcal{W_{\mathrm{cycle}(\mathrm{torus})}}=\prod_{k(,l)=0}^{1}\bigg[\bigotimes_{i\in \mathcal{T}_{k(l)}}^{N-1}W_{i, i+1}(\theta_{i})\bigg]$,
               
               \vspace{2 mm}
    \hrulefill
    
    \textbf{Propagation:}  
    
    \vspace{2 mm}
    
        \For{$(n=1,\ n<N_{stps})$:}{$\ket{\psi_{n}}=\mathcal{W}_{cycle(torus)} \ket{\psi_{n-1}}$ \;} 
    \vspace{2 mm}
        
        \textbf{Return} $\ket{\psi_{N_{stps}}}$ \;
    
    \caption{QW search on N-cycles $\ (N \times N$ torus graphs) via cQED-QCA}\label{algo:algo2}
\end{algorithm}
\begin{algorithm}
    \KwData{N qubits initialized in $\ket{0_{0}, \cdots, 0_{N-1}}$}% $C12's gate-set (Table \ref{tab:0}),}
    %\KwResult{Ideal number of qubit per cavity}
    \hrulefill

        \If{$N > 1$ and N is power of 2}{ 
    \vspace{2 mm}
        
        $\ket{\psi_{0}}=R_{X_{0}}(\frac{\pi}{2})\ket{0_{0}, \cdots, 0_{N-1}}$
        
    $p\_list = [2, 4, \cdots, N/2]$

    \vspace{2 mm}

        \For{$(i=0, \ i<N/2)$:}{
            \If{$i\ \% \ 2 \neq 0 $}{
                \For{k $\mathbf{in}$ $p\_list:$}{
                    \If {$i < k$}{
                        $\ket{\psi_{0}}= R_{Z_{i+k}}(-\frac{\pi}{2}) \cdot U_{i, \ i+k}^{(2)}(\frac{\pi}{4})\ket{\psi_{0}}$ \;}
                        }}
                        }
         %   }
         
       % \For{$(i=0, \ i<N)$:}{
        %    \If{$i\ \% \ 2 \neq 0 $}{
         %   $\ket{\psi_{0}}= R_{Z_{i+2^{N/2}}}(-\frac{\pi}{2}) \cdot U_{i, \ i+2^{N/2}}^{(2)}(\frac{\pi}{4})\ket{\psi_{0}}$ \;}}
    
    \vspace{2 mm}
            
        \For{$(i=0, \ i<N)$:}{
        \If{$i \ \% \ 2 = 0 $}{$\ket{\psi_{0}}=R_{Z_{i}}(-\frac{\pi}{2})\cdot U_{i, \ i+1}^{(2)}(\frac{\pi}{4})\ket{\psi_{0}}$\;
        } 
        }
        }
    \textbf{Return} $\ket{\psi_{0}}$ \;
    
    \caption{\textit{Dicke\_one\_initializer(N)}}\label{algo:algo3}
\end{algorithm}

\textbf{Software}\label{sec:soft}

Algorithms (\ref{algo:algo1},\ref{algo:algo2}) are translated into circuits and executed using Qiskit SDK \cite{Qiskit}. The noiseless simulations are run with Qiskit Aer backend with the following modules and versions: qiskit==0.41.0, qiskit-aer==0.11.2, qiskit-ibmq-provider==0.20.0, qiskit-terra==0.23.1. The noisy simulations are run with \textit{Callisto} $c12-callisto-clients==2.0.1$ \cite{Callisto}, with all-to-all connectivity and the basis gate-set and Fidelity reported in Table (\ref{tab:1}).
%
%\subsection{QW metrics}\label{sec:QW_metrics}
\subsection{QW benchmarks}\label{sec:QW_metrics}
As follows, we outline the quantities and metrics used to analyze the results of the QW and QW search simulations:
\begin{enumerate}
    \item \textit{Qiskit state count distribution}:
    To study the QW (search) algorithm phenomenology, we use Qiskit’s state count function. After executing the walk for a set number of time steps, the register is measured and the walker’s position distribution over the lattice is extracted. This process is repeated across user-defined simulation shots, averaged, and visualized using histogram plots. 
    \item \textit{Hellinger Fidelity}:  To quantify the impact of noise on the simulation, we use the Hellinger Fidelity \cite{Hellinger}, a standard metric for benchmarking quantum algorithms like QW and QW search \cite{Georg1, Cycle, Portugal}. It is defined as: 
\begin{equation}
\begin{split}
    &\mathcal{F_{H}}(P,Q)= \left[1-H(P,Q)^{2}\right]^{2}\\
    &H(P,Q)=\frac{1}{\sqrt{2}} \sqrt{\sum_{i=1}^{K}(\sqrt{p_{i}}-\sqrt{q_{i}})^{2}}\\
\end{split}\label{eq:Hellinger_def}
\end{equation}
where $P$ and $Q$ are probability distributions over $K$ outcomes, and $H(P,Q)$ is the Hellinger distance \cite{Hellinger_dist}. Note that $\mathcal{F_{H}}$ is equivalent to the classical fidelity \cite{Nielsen_Chuang_2010}. In this work, $P$ and $Q$ respectively represent the noiseless (Qiskit Aer) and noisy (\textit{Callisto}) emulator state probability distributions. 
Studying $\mathcal{F_{H}}$ over time steps helps reveal how noise affects QW dynamics. Its values range from 0 (no correlation) to 1 (perfect match). 
While values above $0.5$ are often seen as positively correlated, this can be misleading—e.g., two uniform, uncorrelated distributions may yield high fidelity \cite{Acasiete}. 
Thus, the fidelity analysis should be complemented by visual checks of the state count distribution. In general, $\mathcal{F_{H}} \geq 0.7$ with accurate distributions suggests a successful simulation.

\item $\ell^{1}$ \textit{(Taxicab or Manhattan) distance}: defined as $\ell^{1}(P,Q)=\sum_{i}|p_{i}-q_{i}|$ similarly to the Hellinger Fidelity, this metric quantifies the impact of noise over the QW dynamics by measuring the distance between the noiseless and noisy probability distributions $P$ and $Q$. Such metric is used in \cite{Shakel, Acasiete}.
\end{enumerate}

We now introduce some quantities and metrics specifically used to analyze the QW search\cite{Lovett, Zhang_search}:
\begin{enumerate}

    \item \textit{Hitting time}: it quantifies the number of QW time-steps before the marked vertex probability reaches its maximum. QW search algorithms find the marked vertex in $\mathcal{O}(N)$ in one dimension, $\mathcal{O}(\sqrt{N\log{N}})$ steps in two-dimensions and $\mathcal{O}(\sqrt{N})$ steps for three or more dimensions \cite{Tulsi, Ambainis}. 
    
    \item \textit{Success probability}: it is the maximum probability peak of the marked vertex. The success probability of the QW search scales as $\mathcal{O}(\frac{1}{N})$ in one dimension and $\mathcal{O}(\frac{1}{\log_{2}N})$ in two dimensions \cite{Ambainis, Lovett, Tulsi}.

    \item \textit{Degraded ratio}: it is defined as the ratio between the noiseless (or theoretical) and noisy success probabilities $R=\frac{P_{s, noisy}}{P_{s, ideal}}$. 
    \item \textit{Selectivity}: it is defined as $S=\ln{\frac{P_{v_{m}}}{max\{P_{\Bar{v}}\}}}$ \cite{45,46}, where $P_{v_{m}}$ is the success probability and $max\{P_{\Bar{v}}\}$ the maximum of the non-marked vertex probabilities . The higher the selectivity the more robust to noise a circuit is. 
\end{enumerate}
\subsection{Results}\label{sec:SQWH_sims}
We simulate QW and QW search on the cycle and torus using the cQED QCA models from Section \ref{section:cqed}. Simulations are run with Qiskit SDK \cite{Qiskit} using both the noiseless Aer simulator and C12’s noisy emulator, \textit{Callisto} \cite{Callisto}. Each simulation runs an "experiment" (i.e., Algorithm \ref{algo:algo1} or \ref{algo:algo2} with fixed qubits and time-steps) $10,000$ times (“shots”) to obtain a state distribution.

\subsubsection{Quantum walk}

\textbf{N-cycles}

\begin{figure*}[htp]
  \centering
        \includegraphics[width=6 cm]{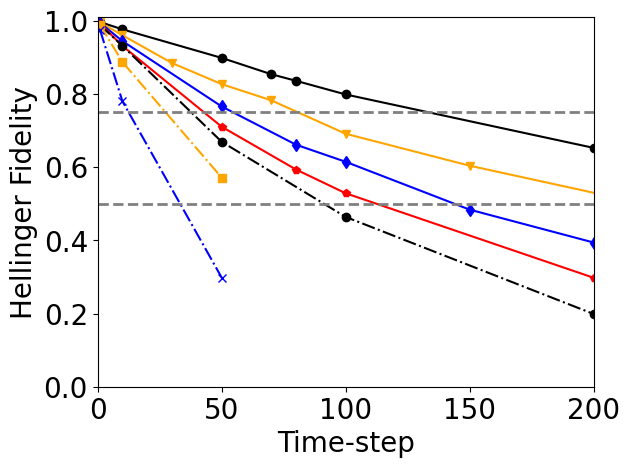}
        \includegraphics[width=6 cm]{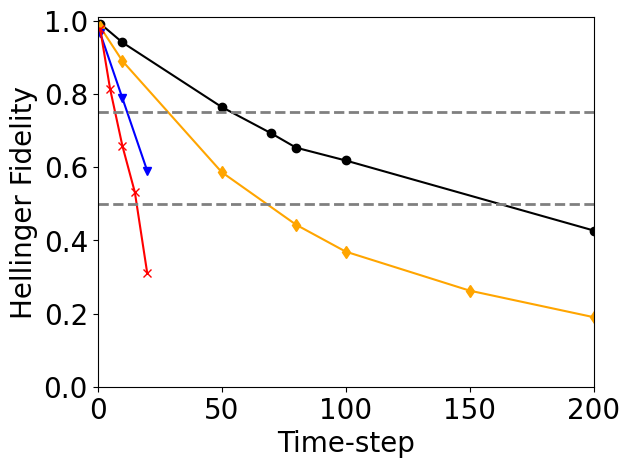}
 \caption{Hellinger Fidelity vs. time-step of the QW (left) and QW search (right) via QCA over 4 (black dots), 6 (orange triangles), 8 (blue diamonds), 10 (red pyramids), 12-cycle (black dashed dot line), 16-cycle (orange dashed squares line) and 4x4-torus (blue dashed cross line).}
 \label{fig:QW_Hell_fidelity}
\end{figure*}

\begin{figure*}[htp]
  \centering
        \includegraphics[width=6 cm]{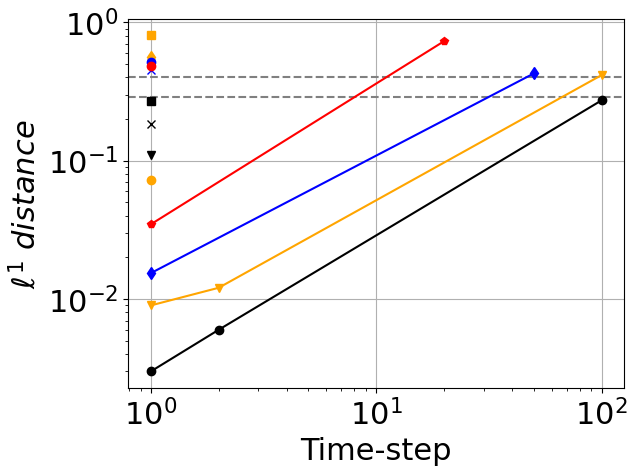}
        \includegraphics[width=6 cm]{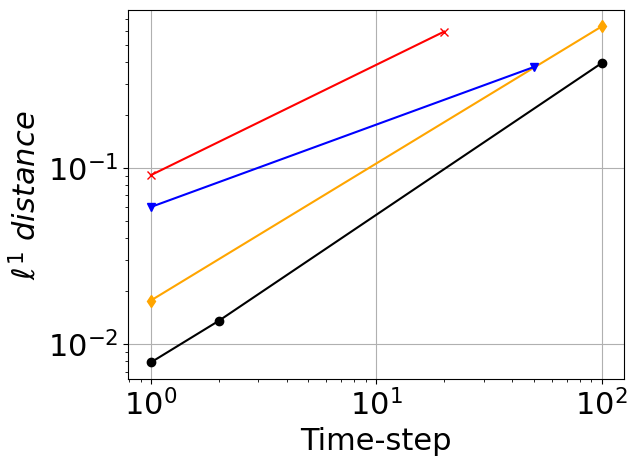}
 \caption{Left: $\ell^{1}$ distance vs. time-step for QW (left) i) via QCA : 4 (black curve), 8 (orange curve), 16-cycles (blue curve) and 4x4 torus (red curve) on \textit{Callisto} ; ii) via \cite{Shakel} QFT: 4 (black: crosses on \textit{ibmqx2}, triangles on \textit{ibmqx2}, squares on \textit{ibmq\_16\_melbourne}), 8-cycles (orange: crosses on \textit{ibmqx2} computer, diamonds on \textit{ibmqx2} processor, squares on \textit{ibmq\_16\_melbourne}) via Shakeel QFT-based QW \cite{Shakel}; iii) via Portugal \cite{Portugal}: 8 (orange dots on \textit{ourense}), 16-cycles (blue crosses on \textit{vigo} processor, diamonds on \textit{ibmqx2} and 4x4 torus (red dots on \textit{vigo}). Right: $\ell^{1}$ distance vs. time-steps for QW search (right) i) via QCA : 4 (black), 8 (orange), 16-cycles (blue) and 4x4 torus (red) on \textit{Callisto}.}
 \label{fig:l1_distance}
\end{figure*}

We simulate QWs on cycles of size 4, 6, 8, 10, 12 and 16 using our cQED QCA implementation. To validate the dynamics, we analyze state count distributions over time; an example for the 8-cycle is shown in Fig. (\ref{fig:8_cycle_QW_histo}), displaying typical QW behavior, consistent with QW patterns \cite{Kitagawa:2010xsb, Acasiete}. Noiseless and noisy results align closely in early steps.
To quantify noise effects, we track the Hellinger Fidelity and the $\ell^{1}$ distance over time-steps Fig. (\ref{fig:QW_Hell_fidelity},\ref{fig:l1_distance}). The Fidelity decreases with time due to cumulative gate errors and more rapidly with larger cycles, as the gate count scales with system size (Section \ref{section:cqed}).
For a very high number of time-steps, we expect all the curves to reach an asymptote, corresponding to a completely random output due to noise.
Our results compare well with prior work \cite{Acasiete, Cycle, Georg1}. 
In \cite{Acasiete}, the authors report: (i) For a 16-cycle QW, after one step, fidelities of $ \approx 0.7170$ with \textit{ibmqx2} quantum computer and $\approx 0.7358$ with \textit{vigo} \cite{ibm_exp}; by contrast, our simulation achieves $\approx 0.6380$ after fifty steps (Fig. \ref{fig:QW_Hell_fidelity}).
(ii) For an 8-cycle, \textit{Ourense} reaches $\approx 0.90$ after eight steps, though the corresponding state-count distribution shows the noisy output to be wrong. In our case, the fidelity reaches the same value at step 18, with state count distributions showing good agreement between noisy and noiseless outputs.
Similar considerations hold for the $\ell^{1}$ distance analyses. The bottom plot of Fig. (\ref{fig:l1_distance}), reports $\ell^{1}$ against the QW time-step for different cycle size, circuit implementations (via QCA, \cite{Shakel} and \cite{Portugal}) and quantum computer or emulators. The QCA implementation presented in this work scores a consistently lower $\ell^{1}$ distance with respect to other implementations, i.e. around one order of magnitude lower for the same time-step and one order of magnitude later (in time-steps) for the same $\ell^{1}$ value. The positive performance of our QCA implementation highlights its potential: despite higher qubit requirements, it benefits from efficient linear gate scaling. 
Obviously, we stress that our results are based on emulator runs, unlike the hardware-based results used for comparison, and thus further testing on advanced emulators or real quantum hardware is necessary to fully assess the impact of noise over our QCA implementations.
\begin{figure*}[htp]
  \centering\begin{tabular}{
  @{}c@{ }c@{ }c@{ }c@{}}
&\textbf{\Large4-cycle} & \textbf{\Large8-cycle} & \textbf{\Large16-cycle} \\
        &\includegraphics[width=5.5 cm]{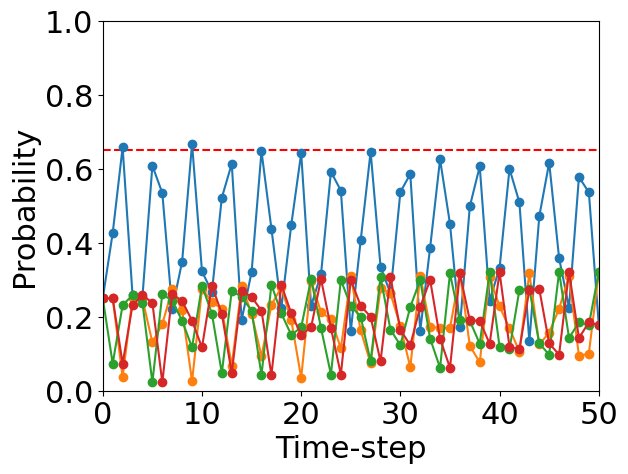}&
        \includegraphics[width=5.5 cm]{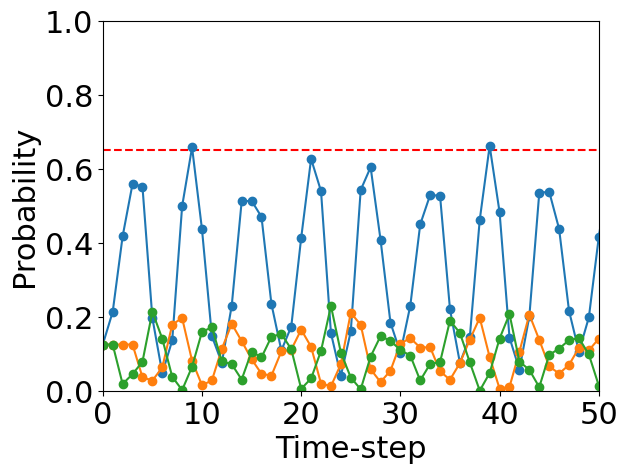}&
        \includegraphics[width=5.5 cm]{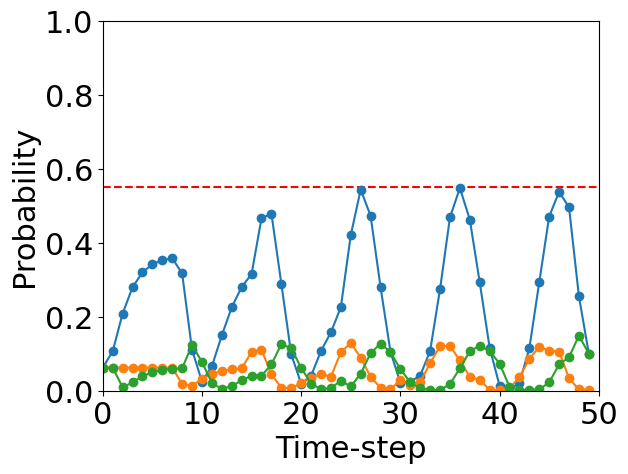}\\
        
        &\includegraphics[width=5.5 cm]{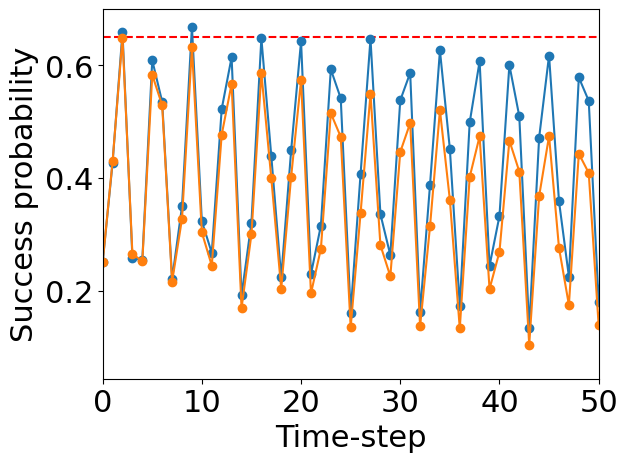}&
        \includegraphics[width=5.5 cm]{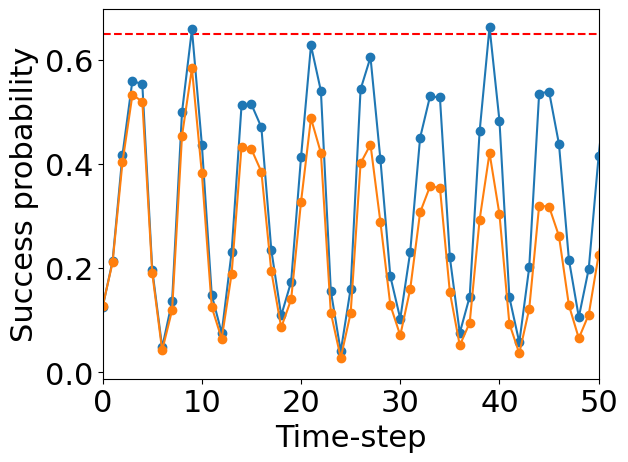}&
        \includegraphics[width=5.5 cm]{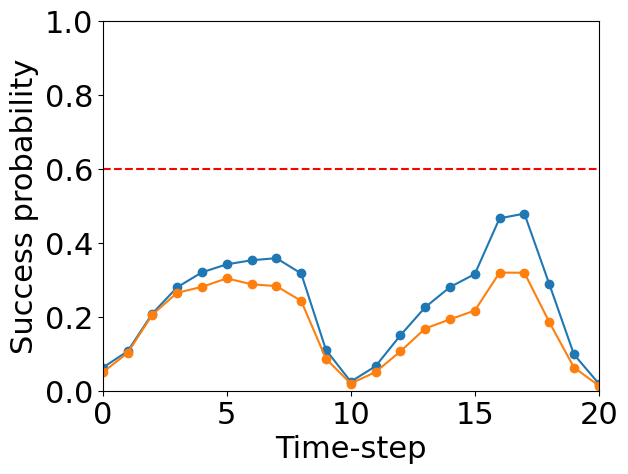}\\
        
        &\includegraphics[width=5.5 cm]{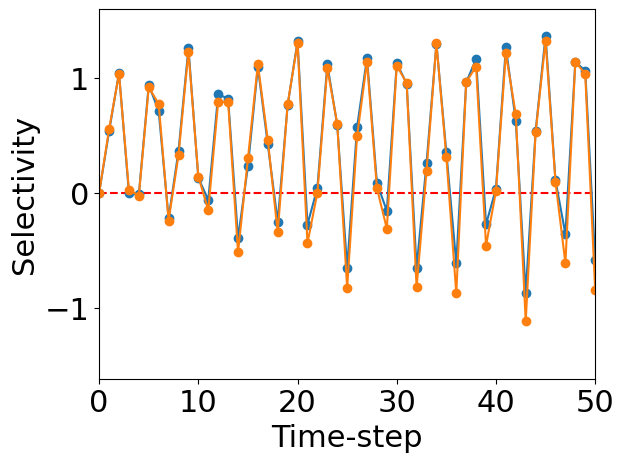}&
        \includegraphics[width=5.5 cm]{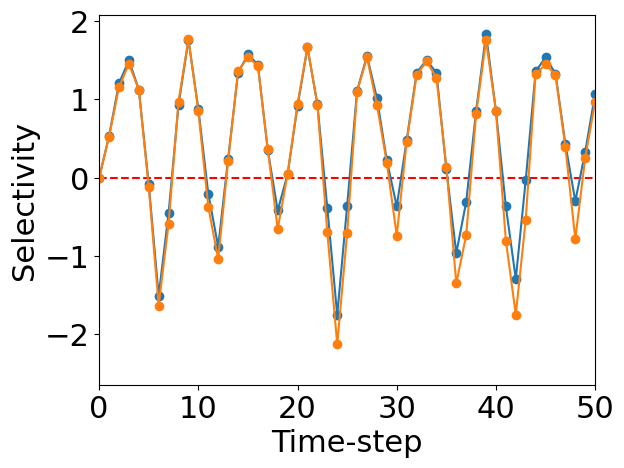}&
        \includegraphics[width=5.5 cm]{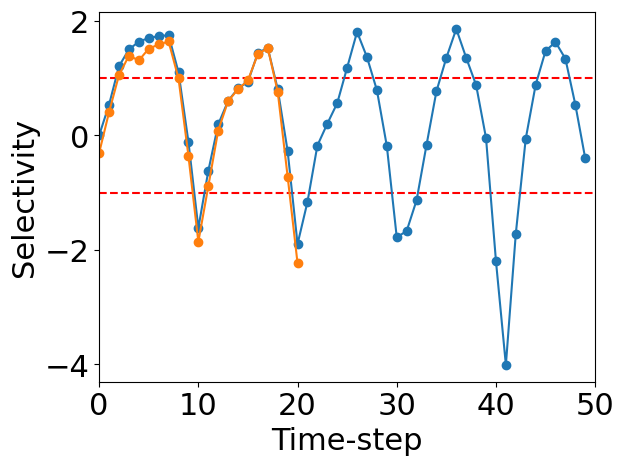}\\  
\end{tabular}
 \caption{Noiseless (Aer) and noisy (\textit{Callisto}) simulations of QW search via QCA. First row: noiseless vertex probabilities vs. time-step (for 8 and 16-cycle some vertex curves are not shown for visual clarity). Second row: noiseless (orange) and noisy (blue) success probabilities vs. time-step. Third row: noiseless (orange) and noisy (blue) selectivity vs. time-step.}
 \label{fig:QW_metrics_1D}

\end{figure*}
\begin{figure*}[htp]
  \centering 
         \includegraphics[width=5.5 cm]{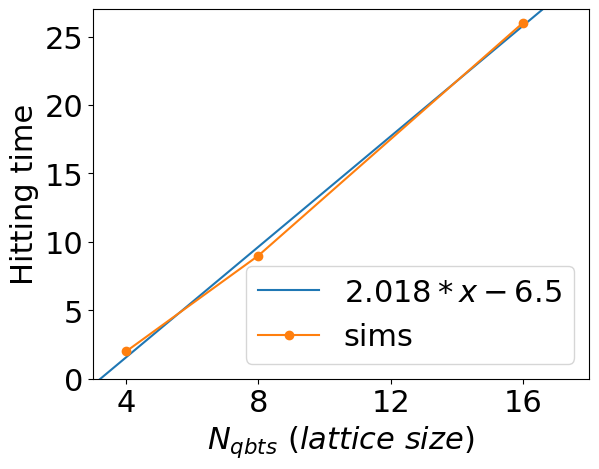}
        \includegraphics[width=5.5 cm]{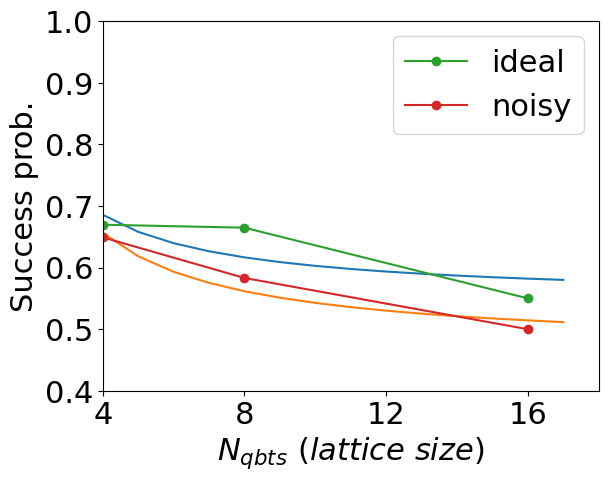}
        \includegraphics[width=5.9 cm]{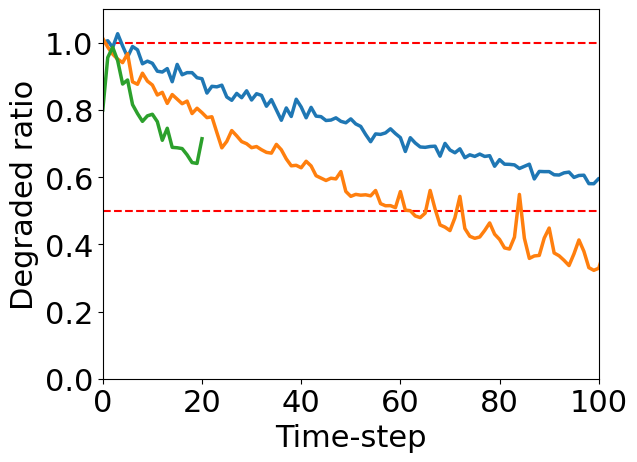}
 \caption{Noiseless (Aer) and noisy (\textit{Callisto}) simulations of QW search via QCA over 4, 8 and 16-cycle. Left: Hitting time vs. number of qubits (lattice size) along with a linear fit. Middle: noiseless and noisy success probabilities vs. number of qubits (lattice size) along with a inverse linear fit. Right: degraded ratio vs. time-step.}
 \label{fig:QW_metrics_1D}
\end{figure*}
%\end{widetext}
%

\textbf{$4\times 4$ torus}

We simulate QW on a $4\times4$ torus. The state count histograms (Fig. \ref{fig:4x4_torus_QW_search_2Dhisto}, in App. \ref{sec:app}) show behavior dependent on the initial condition, with good alignment between noisy and noiseless outputs in early steps. To quantify noise effects, we analyze the Hellinger Fidelity and the $\ell^{1}$ distance over time (Fig. \ref{fig:QW_Hell_fidelity},\ref{fig:l1_distance}). Compared to 1D cycles, 2D metrics degrade faster due to a doubled number of two-qubit gates per step, as showed in Eq. (\ref{eq:QCA_tr_func}).
Compared to \cite{Portugal}, which reports a Fidelity of $\approx 0.51$ and $\ell^{1} \approx 0.485$
after one step, our implementation achieves 0.92 and 0.035, respectively, reaching their values only after $\approx 10 \times$ more steps. Once again, we stress that these results are from an emulator, and further testing on real hardware or advanced emulators is needed to fully gauge noise impact.
\begin{figure*}[htp]
  \centering
        \includegraphics[width=6 cm]{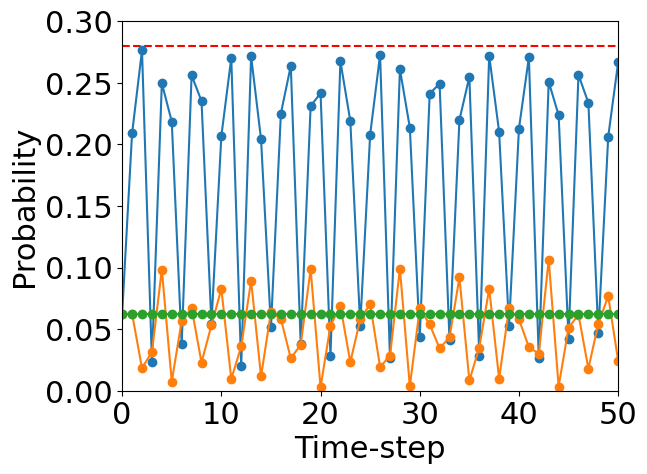}
         \includegraphics[width=6 cm]{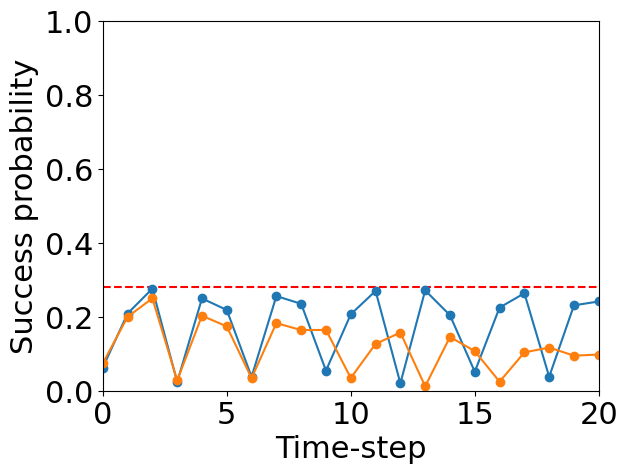}
          
          \includegraphics[width=6 cm]{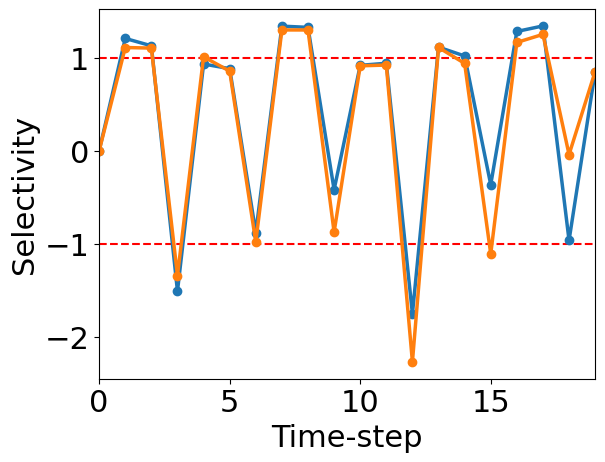}
           \includegraphics[width=6 cm]{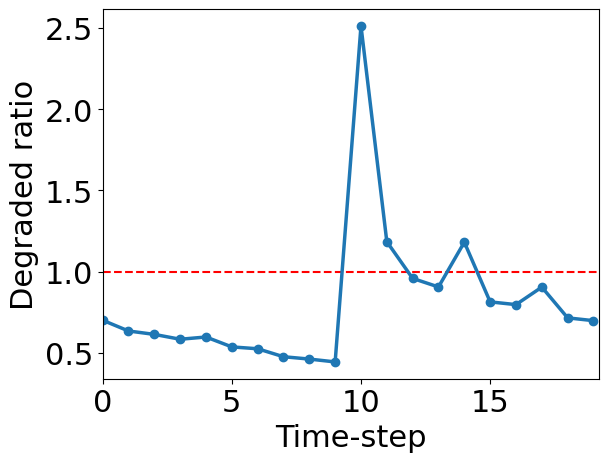}
 \caption{Noiseless (Aer) and noisy (\textit{Callisto}) simulations of QW search via QCA over $4\times4$ torus. Top left: noiseless vertex probabilities vs. time-step (overlapping vertex curves are not shown for visual clarity). Top right: noiseless (orange) and noisy (blue) success probabilities vs. time-step. Bottom left: noiseless (orange) and noisy (blue) selectivity vs. time-step. Bottom right: degraded ratio vs. time-step.}
 \label{fig:QW_metrics_2D}
\end{figure*}
\subsubsection{Quantum Walk search}

\textbf{4, 8 and 16-cycles}

We simulate QW search on 4, 8, and 16-cycles. State count histograms (Fig. \ref{fig:8_cycle_QW_search_histo}, in App. \ref{sec:app}) show correct behavior, with higher probability at the marked vertex.
To further assess the performance of our QCA implementation, we use the metrics in Sec. \ref{sec:QW_metrics} (see Fig. \ref{fig:QW_metrics_1D}):

\begin{enumerate}
    
    \item \textbf{Marked vertex probability:} consistently higher than the non-marked vertex probabilities, modulo some oscillations due to the lattice finite size effects, (see first and second row of Fig. \ref{fig:QW_metrics_1D}). Selectivity plots (third row of Fig. \ref{fig:QW_metrics_1D}) confirm that larger lattices yield more stable amplification. The similarity between the noisy (orange) and noiseless (blue) curves confirms that the oscillations stem from finite lattice-size effects. Indeed, larger lattices result in a more stable marked vertex probability, as shown in the torus selectivity study (Fig. \ref{fig:QW_metrics_2D}).
    \item \textbf{Hitting time:} display a trend with acceptable agreement with a linear fit (first plot, last row of Fig. \ref{fig:QW_metrics_1D}), consistent with theoretical predictions \cite{Lovett}, although additional data is required to confirm its validity.
    \item \textbf{Success probability} decreases with the lattice size (second plot of last row of Fig. \ref{fig:QW_metrics_1D}), following a trend which could be interpreted as linearly inverse with the lattice size (number of qubits). %For 4, 8, and 16-cycles, we observe success probabilities of approximately 0.65, 0.60, and 0.55, respectively. 
    Our \textit{scipy} fit supports this, but more data is needed to accurately characterize the trend.
    \item \textbf{Degraded ratio:} $R_{s}=\frac{P_{s, noisy}}{P_{s, ideal}}$ (Fig. \ref{fig:QW_metrics_1D}, last plot of last row), shows that noise reduces probabilities uniformly across marked and non-marked vertices, preserving search functionality.

    \item \textbf{Hellinger Fidelity and $\ell^{1}$ distance:} Both metrics (Fig. \ref{fig:QW_Hell_fidelity},\ref{fig:l1_distance}) degrade faster for larger lattices and for QW search (vs standard QW) due to more expensive initialization  and use of lower-fidelity gates (see Table \ref{tab:1} and Algorithms \ref{algo:algo1},\ref{algo:algo2}). As time-steps increase, the metrics approach values expected from random output.   
\end{enumerate}
Once again, we emphasize that our simulations are run on an emulator, not a real quantum computer, so further testing on advanced emulators or actual hardware is needed to accurately assess noise impact.

\textbf{$4 \times 4$ torus}

We simulate QW search on a $4\times4$ torus. The state count histograms (Fig. \ref{fig:4x4_torus_QW_search_histo}, \ref{fig:4x4_torus_QW_search_2Dhisto}) show correct QW search behavior, with minor discrepancies between noiseless and noisy counts in early steps. 
Benchmarking metrics and quantities (Fig. \ref{fig:QW_metrics_2D}) reveal a success probability of $\approx 0.28$ - comprehensibly lower than the one-dimensional success probabilities - being hit after two time-steps and noise-induced oscillations in the degraded ratio.  The Hellinger Fidelity and $\ell^{1}$ distance (Fig. \ref{fig:QW_Hell_fidelity},\ref{fig:l1_distance}) show faster degradation in 2D compared to 1D, due to the increased number of two-qubit gates per time step on the torus (see Eq. \ref{eq:QCA_tr_func}).
To our knowledge, this is the first evaluation of QW search on a torus using QCA on a NISQ processor. Further simulations on larger 2D or higher-dimensional lattices are needed to better assess the advantages and limitations of QW search via QCA.
\section{Conclusion}\label{section:concl}
In this work, we mapped QCA implementations of QW and QW search algorithms over N-cycles and $N \times N$ torus graphs  to C12's Quantum Electronics cQED hardware. Using Qiskit SDK, we ran noiseless and noisy simulations, respectively with Qiskit Aer simulator and C12's \textit{in-house} emulator \textit{Callisto}.
In particular, we ran simulations over different cycle sizes and on a $4\times 4$ torus. We analyzed the simulation outputs with Qiskit's state count histogram plots (App. \ref{sec:app}) , which, serving as a proof-of-concept, show our implementation to correctly reproduce the expected QW and QW search behavior. %In particular, the QCA model we used generates a QW dynamic similar to the Staggered QW \cite{Portugal_stag, Acasiete}. This choice is motivated by the fact that this QCA efficiently minimizes the number of qubits and two-qubit gates required per time step.

For both the QW and QW search, we assessed the noise impact on our simulation results using the Hellinger Fidelity and $\ell^{1}$ distance, benchmarking our implementation. Our study shows, see Fig. \ref{fig:QW_Hell_fidelity},\ref{fig:l1_distance}, that the two metrics gradually deteriorate with the number of algorithm steps (or, equivalently, circuit depth), due to gate errors stacking up at each time-step, thus causing leakage to incorrect states, whether on or off the lattice sites. 
For the QW, a comparison of our QCA implementation over the emulator \textit{Callisto} with similar studies \cite{Acasiete, Shakel} shows promising results. However, being our simulations ran on an emulator, further investigations with a more advanced emulator or a real quantum processor are needed to better understand noise effects.
For the QW search instead, we ran simulations with the emulator Callisto on 4-, 8-, 16-cycles and 4×4 torus graphs, and analyzed the results using suitable metrics and quantities, such as hitting time, success probability, selectivity, and degraded ratio \cite{Zhang_search2} (see Fig. \ref{fig:QW_metrics_1D}). In 1D, we compared our results with theoretical predictions for hitting time and success probability. Selectivity showed oscillations in the marked vertex probability due to finite-size effects in both 1D and 2D. Noise analysis, using the degraded ratio, Hellinger fidelity, and $\ell^{1}$ distance, shows stability within the first hundred time steps, with larger lattices experiencing more noise. Further simulations on larger and higher-dimensional lattices are needed for a deeper understanding of QW search via QCA.

To the authors' knowledge, this work provides the first investigation of QW and QW search implementations via QCA on a NISQ processor emulator. 
Our analysis benchmarks C12's quantum computer via its emulator, \textit{Callisto}, and the QCA algorithmic framework, demonstrating their effectiveness in early NISQ implementations of QW and QW search algorithms. The non-interacting one-particle sector QCA model, equivalent to the QW, can be easily implemented on C12's hardware, requiring for its time evolution operator the XY-gate present in C12's basis gate-set.
More broadly, our study shows the QCA fruitfulness as an algorithmic framework for NISQ implementations, due to its \textit{close-to-physics} spatio-temporal definition. Indeed, despite an unfavorable linear scaling of the lattice size with the number of qubits, the cycle and torus QW and QW search implementations via QCA, require a time-evolution operator composed of common two-qubit gates only. This translates into a favorable linear scaling of the two-qubit gate-count and circuit's depth with the algorithms steps, scarce of big overhead due to complex gate decomposition proper of other circuit implementations. %

While the emulator we employed accounts for the main sources of error in the processor, primarily those arising from incoherent processes, it could be further refined by incorporating coherent errors, such as quantum cross-talk, which have already been investigated in these platforms \cite{mammola}. Although cQED spin-qubit platforms are not yet readily available for experimental use, following the first demonstration of a two-qubit gate in 2024 \cite{dijkema2025cavity}, their availability for experiments can reasonably be expected within the next two to three years.

In conclusion, our results are promising, yet further investigation is required for testing our implementations: i) over different lattices or higher-dimensional structures \cite{Childs_exp_speedup, Childs_forget}; ii) over some use-cases, like asset-price evolution modeling in finance \cite{DEBACKER2025130215} or for network analyses \cite{Chawla2019DiscretetimeQW}; iii) for state preparation, e.g. in  \cite{Chang:2023nls}. Additionally, exploring more complex simulations, such as QFT theories or topological phases of matter \cite{Farrelly, Rydberg}, with initial tests on the Schwinger model (1+1)-QED \cite{Schwinger}, would be valuable.
\begin{acknowledgments}
This work is supported by the PEPR integrated project EPiQ ANR-22-PETQ-0007; by the ANR JCJC DisQC ANR- 22-CE47-0002-01 founded from the French National Re- search Agency; and the French government under the France 2030 investment plan, as part of the Initiative d’Excellence d’Aix-Marseille Université—A*MIDEX AMX-21-RID-011.
We thank Viktor Radovic for his great help with simulations and insightful discussions and Mathieu Roget for the reviewing of the manuscript.
\end{acknowledgments}
%
%\clearpage
%

\appendix
\section{Simulations plots}\label{sec:app}
\begin{figure}[htp]
  \centering
        \includegraphics[width=4 cm]{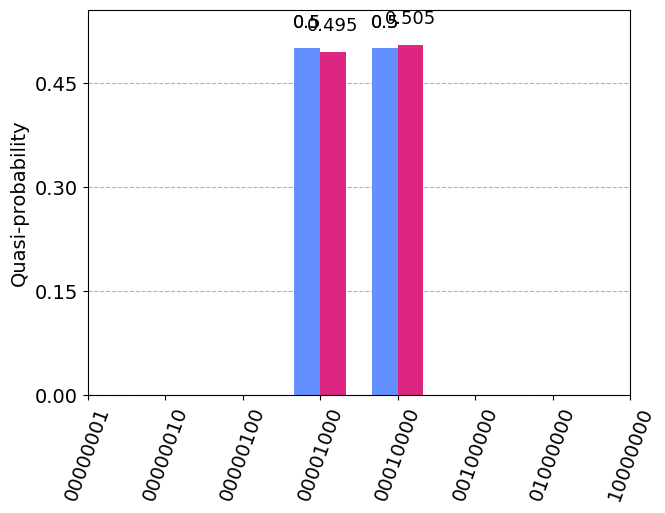}
        \includegraphics[width=4 cm]{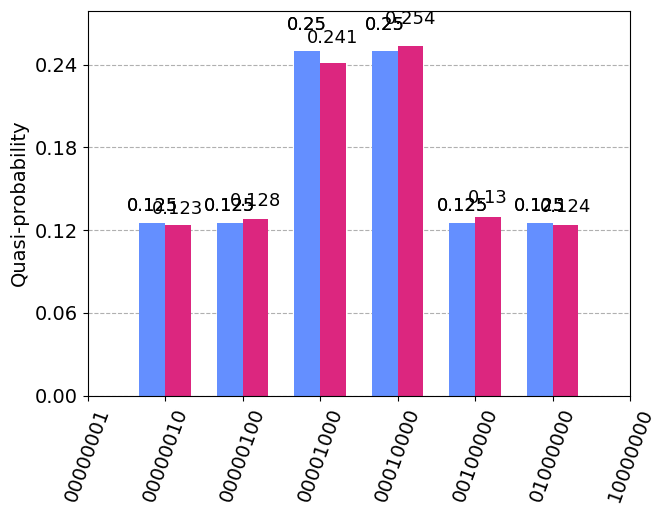}
        
        \includegraphics[width=4 cm]{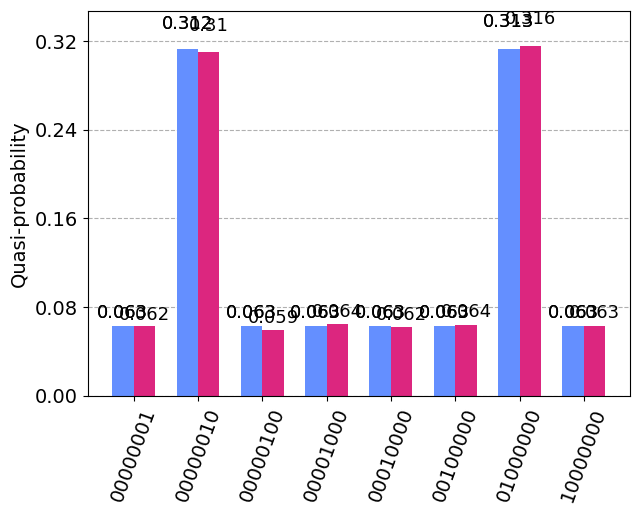}
        \includegraphics[width=4 cm]{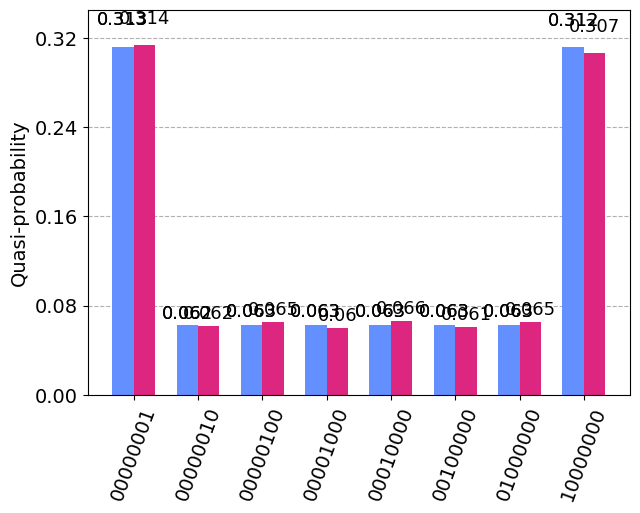}
        
        \includegraphics[width=4 cm]{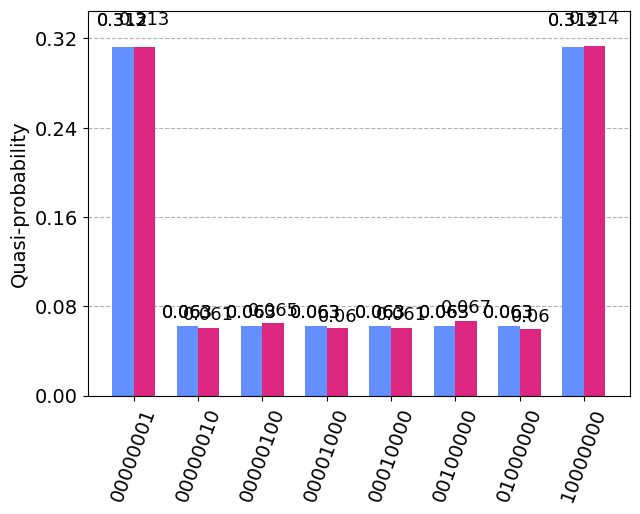}
        \includegraphics[width=4 cm]{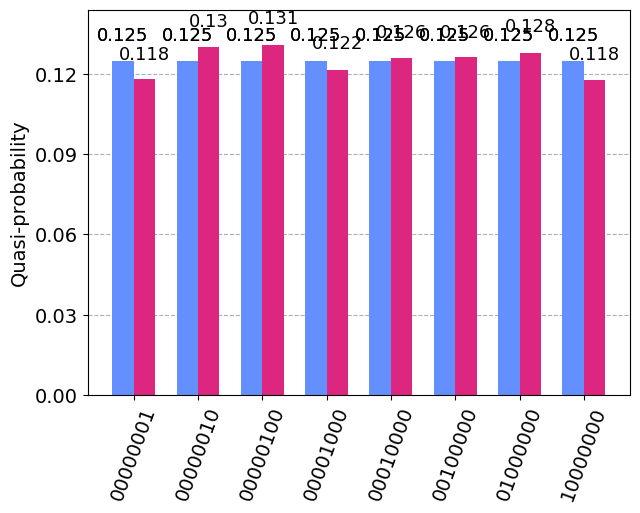}
        
        \includegraphics[width=4 cm]{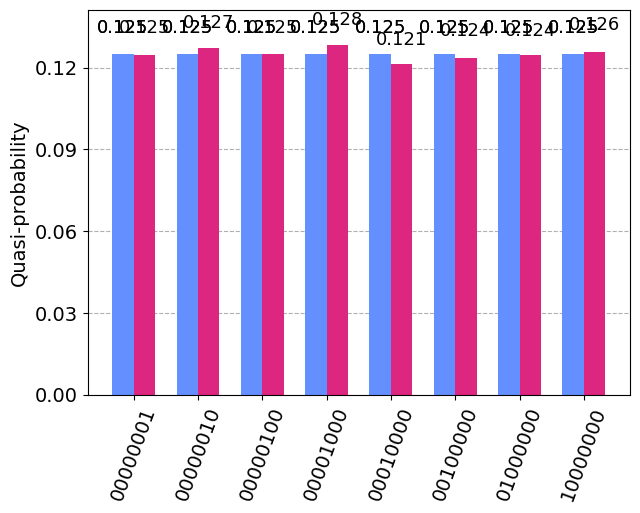}
        \includegraphics[width=4 cm]{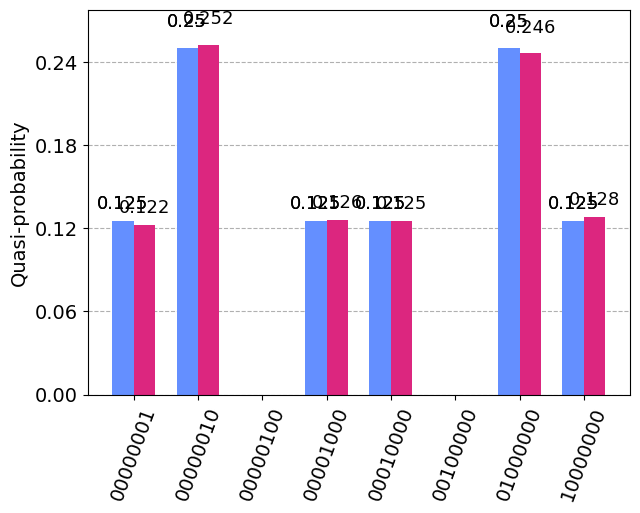}
        
        \includegraphics[width=4 cm]{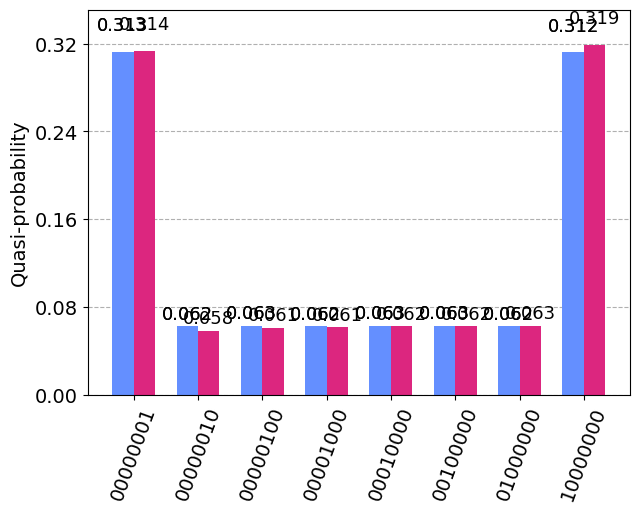} 
        \includegraphics[width=4 cm]{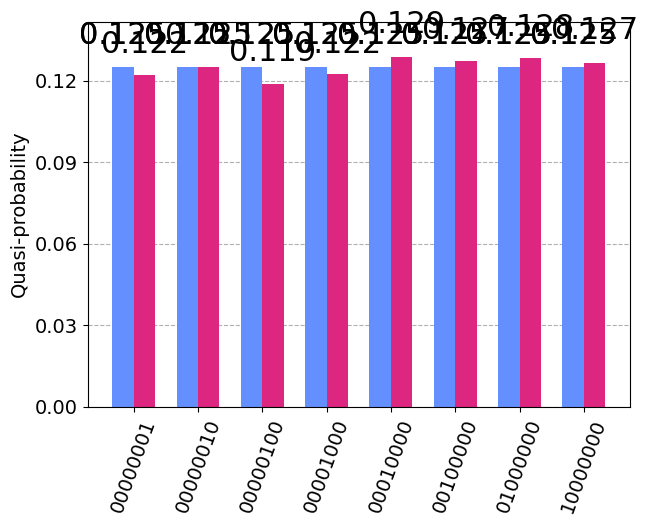} 
 \caption{8 steps 8-cycle noiseless (blue) and noisy (purple) with 10 000 shots, QW via QCA for initial state $\ket{\psi_{0}}=\frac{1}{\sqrt{2}}\ket{000}(\ket{3}+\ket{4})\ket{000}$. Last bottom right plot shows the 18th step.}
 \label{fig:8_cycle_QW_histo}
\end{figure}
\begin{figure}[htp]
  \centering
        \includegraphics[width=4 cm]{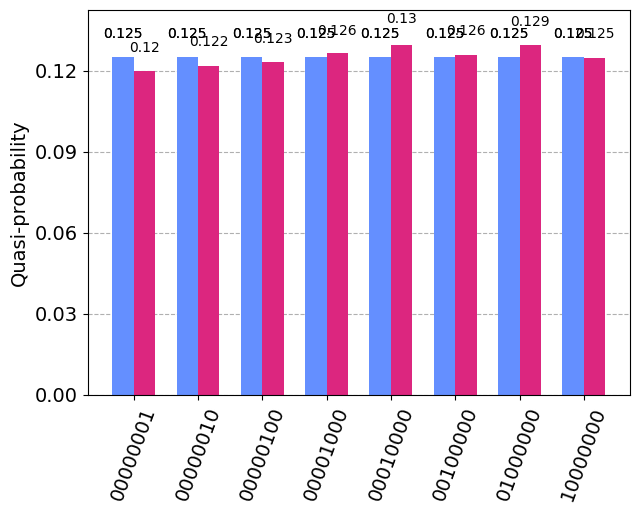}
        \includegraphics[width=4 cm]{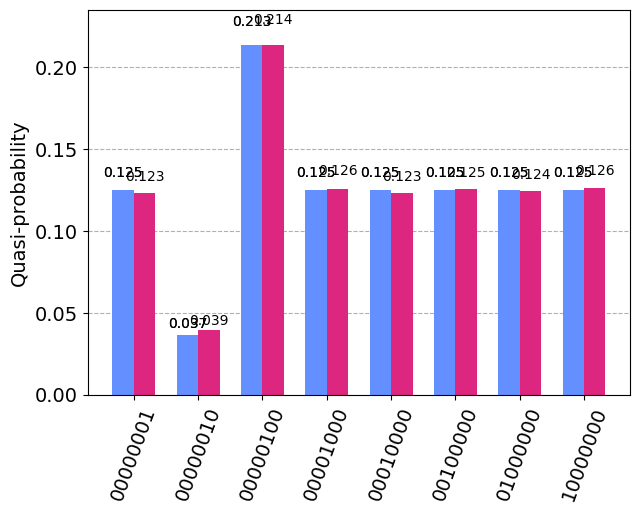}
        \includegraphics[width=4 cm]{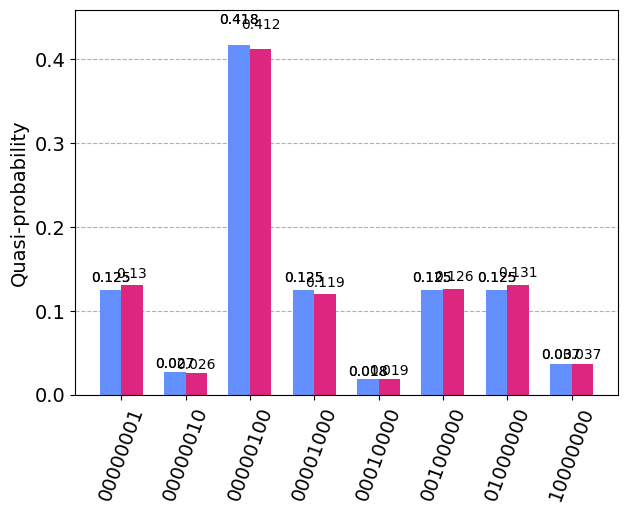}
        \includegraphics[width=4 cm]{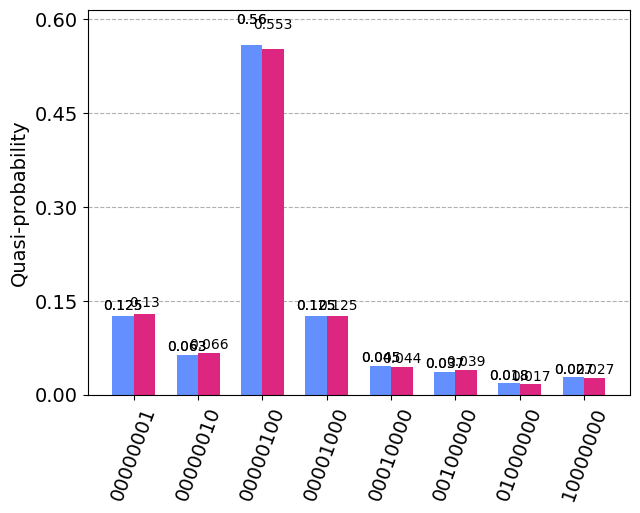}
        \includegraphics[width=4 cm]{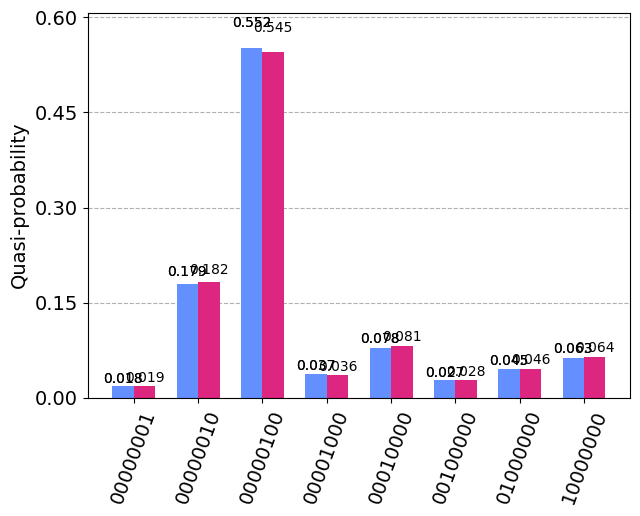}
        \includegraphics[width=4 cm]{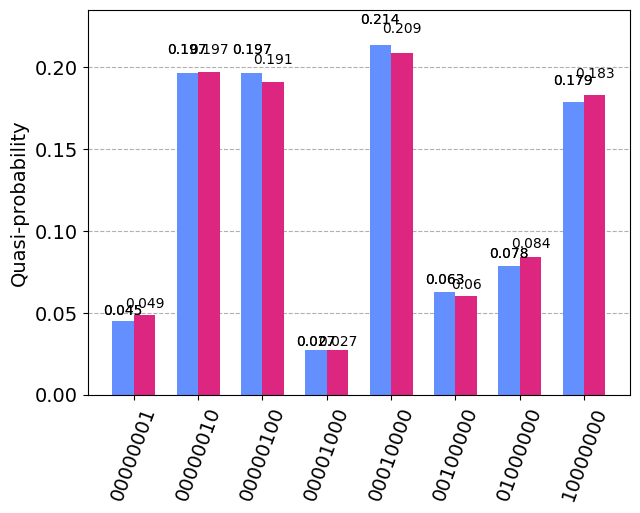}
        \includegraphics[width=4 cm]{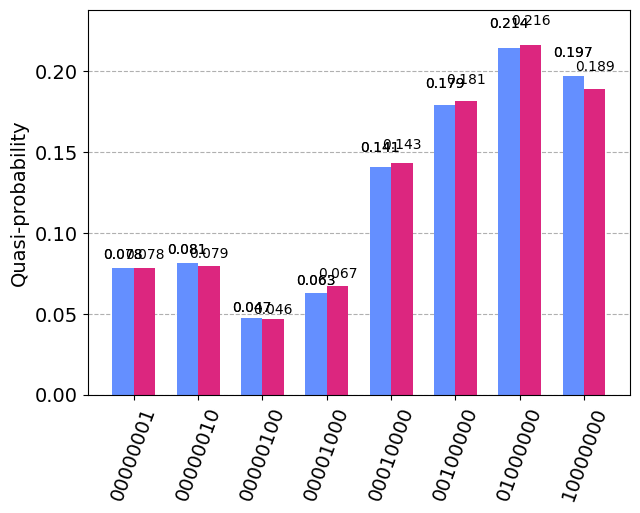}
        \includegraphics[width=4 cm]{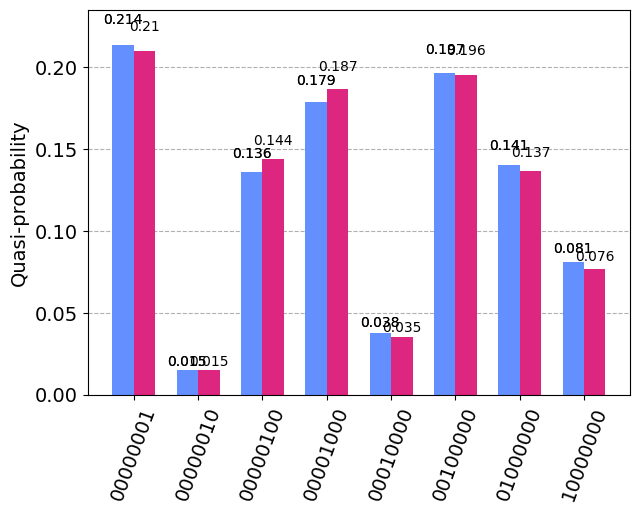}
        \includegraphics[width=4 cm]{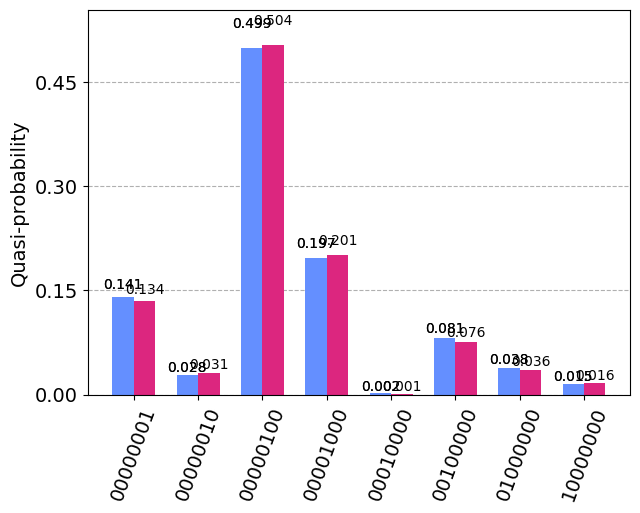}

 \caption{8 steps of QW search via QCA over an 8-cycle with marked vertex equal to 2. Noiseless Qiskit Aer (blue) and noisy \textit{Callisto} (purple) simulations with 10000 shots.}
 \label{fig:8_cycle_QW_search_histo}
\end{figure}
\begin{figure*}[htp]
  \centering
        \includegraphics[width=4.5 cm]{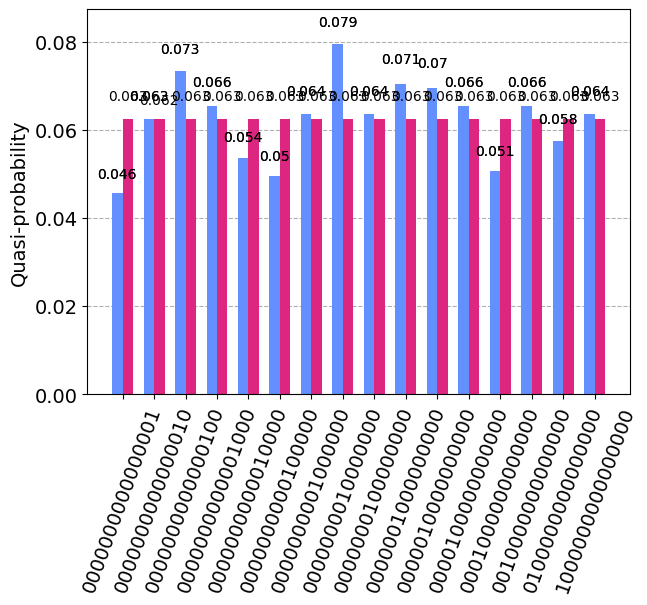}
        \includegraphics[width=4.5 cm]{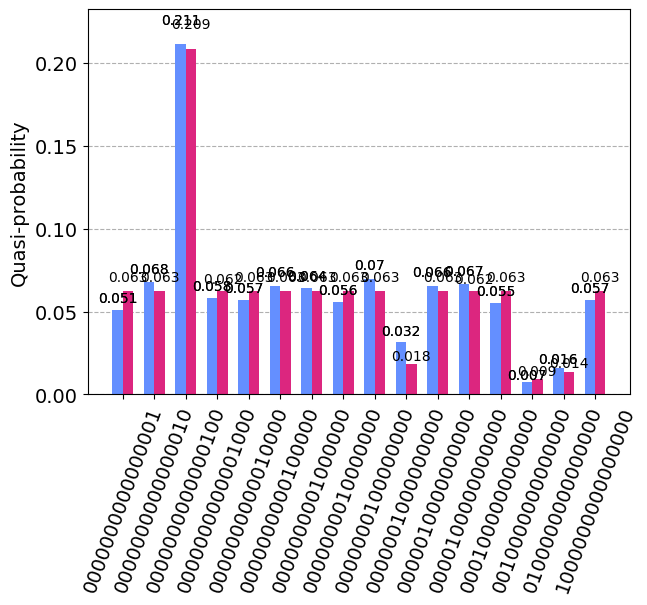}
        \includegraphics[width=4.5 cm]{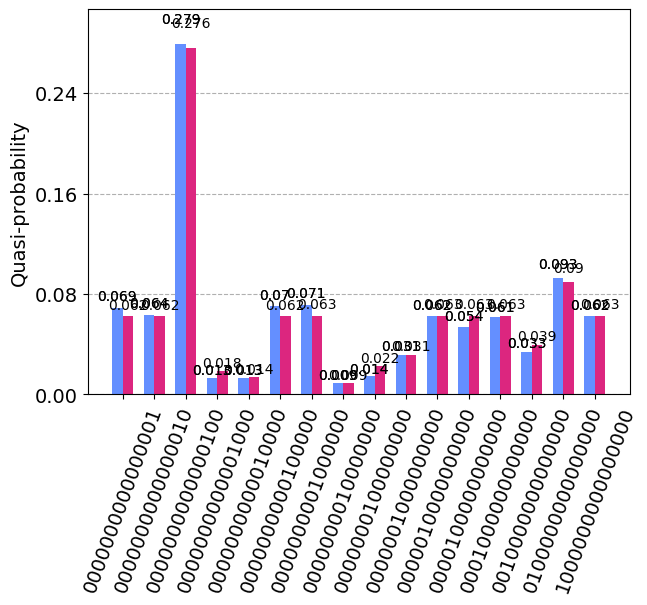}

        \includegraphics[width=4.5 cm]{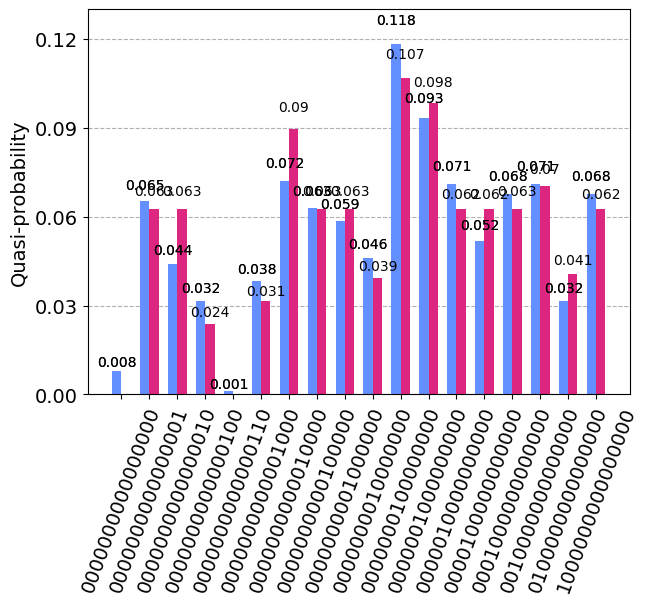}
        \includegraphics[width=4.5 cm]{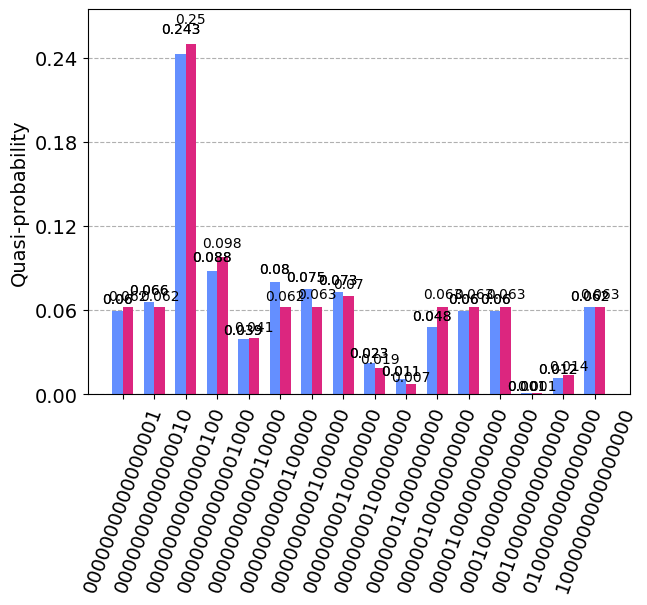}
        \includegraphics[width=4.5 cm]{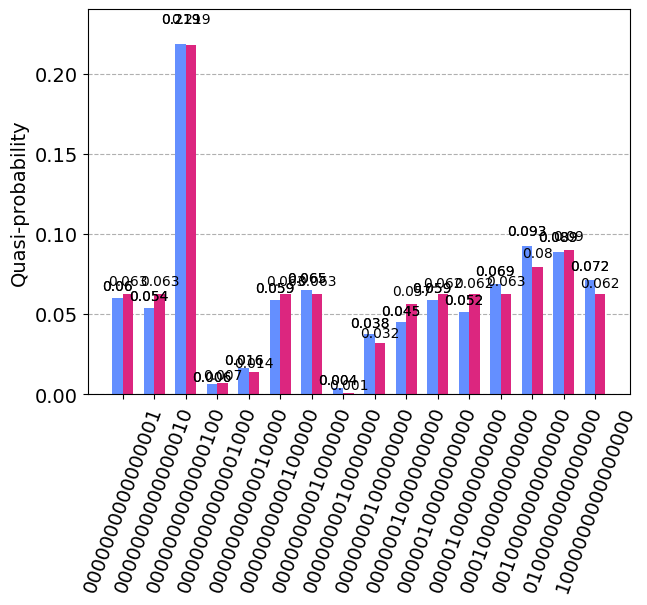}

        \includegraphics[width=4.5 cm]{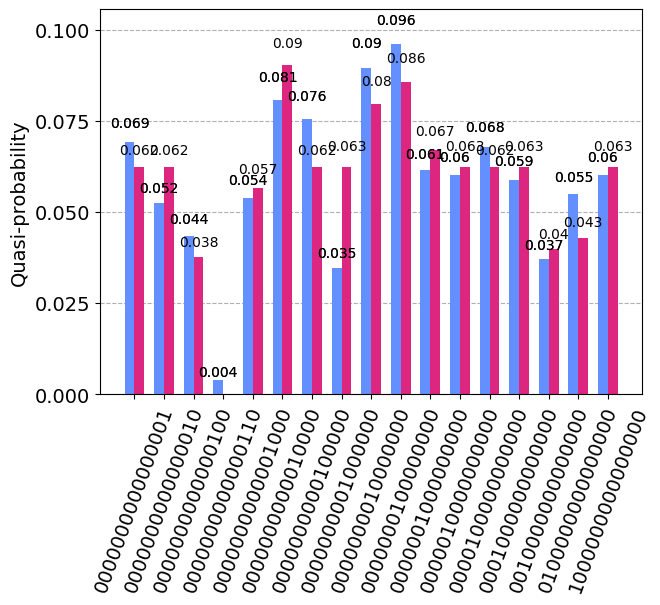}
        \includegraphics[width=4.5 cm]{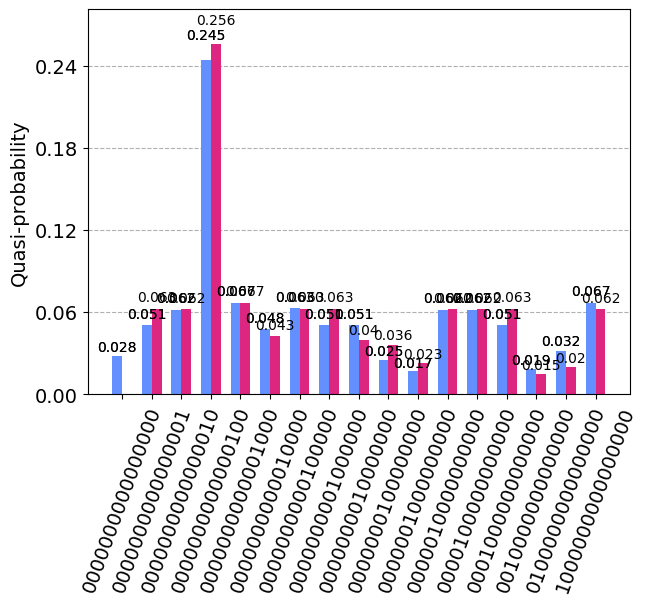}
        \includegraphics[width=4.5 cm]{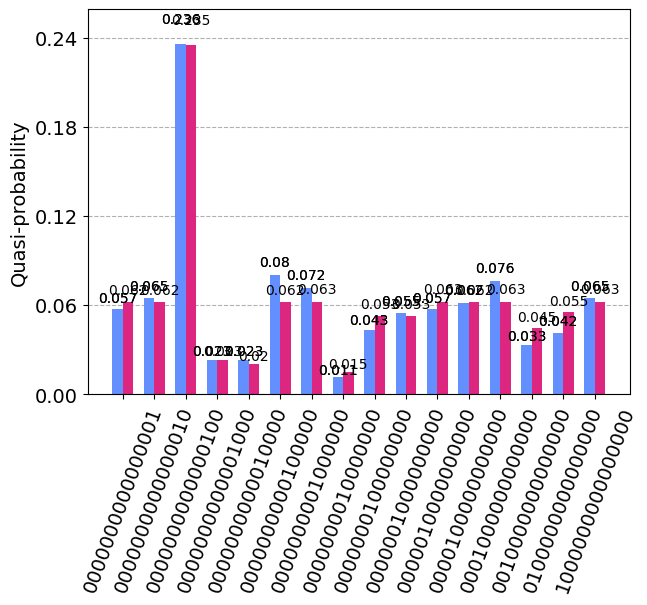}
 \caption{8 steps of QW search via QCA over a 4x4-torus with marked vertex (3,0). Noiseless Qiskit Aer (blue) and noisy \textit{Callisto} (purple) simulations with 10000 shots.} \label{fig:4x4_torus_QW_search_histo}
\end{figure*}
\begin{figure*}[htp]
  \centering
        \includegraphics[width=4 cm]{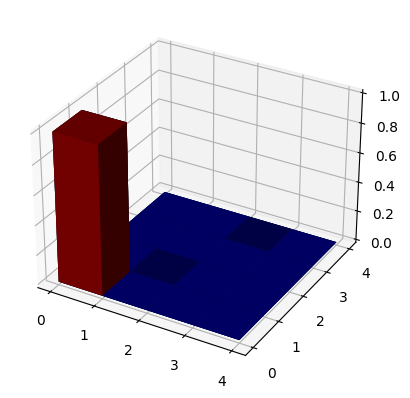}
        \includegraphics[width=4 cm]{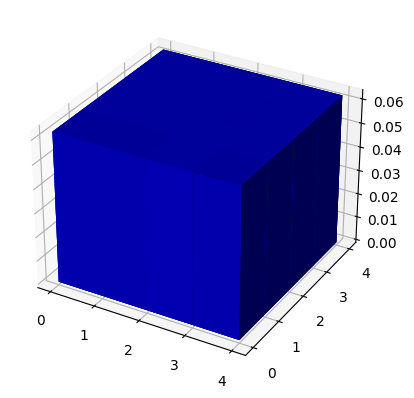}
        \includegraphics[width=4 cm]{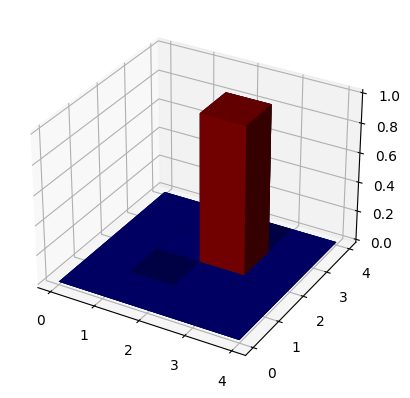}
        
        \includegraphics[width=4 cm]{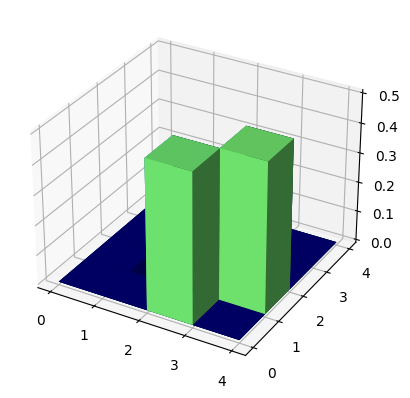}
        \includegraphics[width=4 cm]{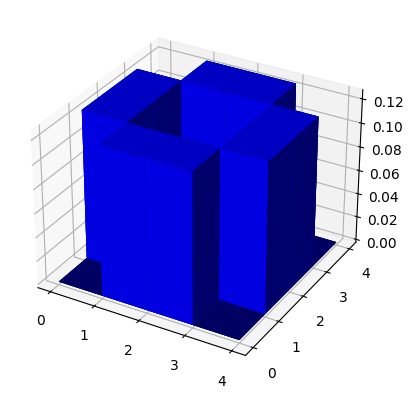}
        \includegraphics[width=4 cm]{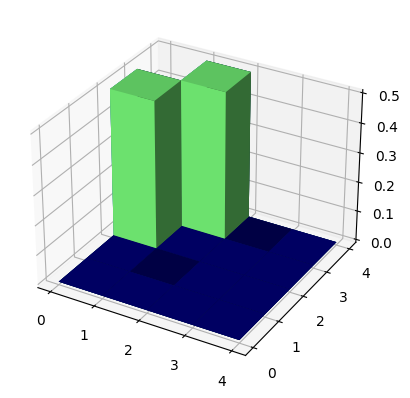}

        \includegraphics[width=4 cm]{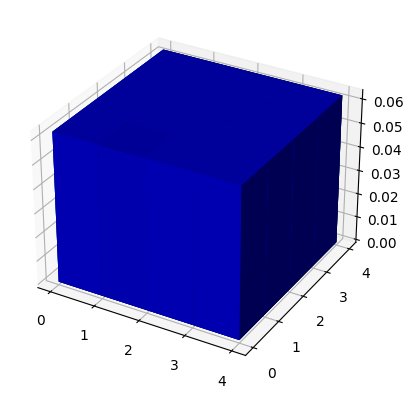}
        \includegraphics[width=4 cm]{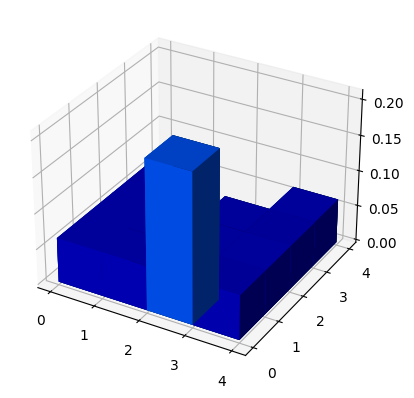}
        \includegraphics[width=4 cm]{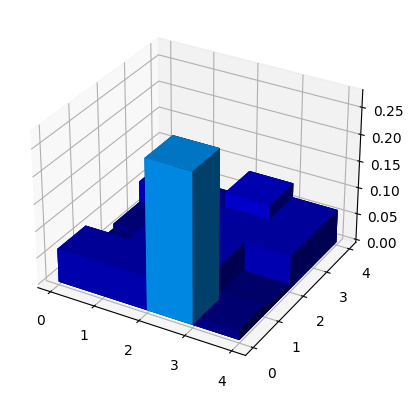}
 \caption{3 steps of QW (top, middle) and QW search (bottom) over a 4x4-torus via QCA. In the QW search the marked vertex is (3,0).} \label{fig:4x4_torus_QW_search_2Dhisto}
\end{figure*}
\begin{figure*}
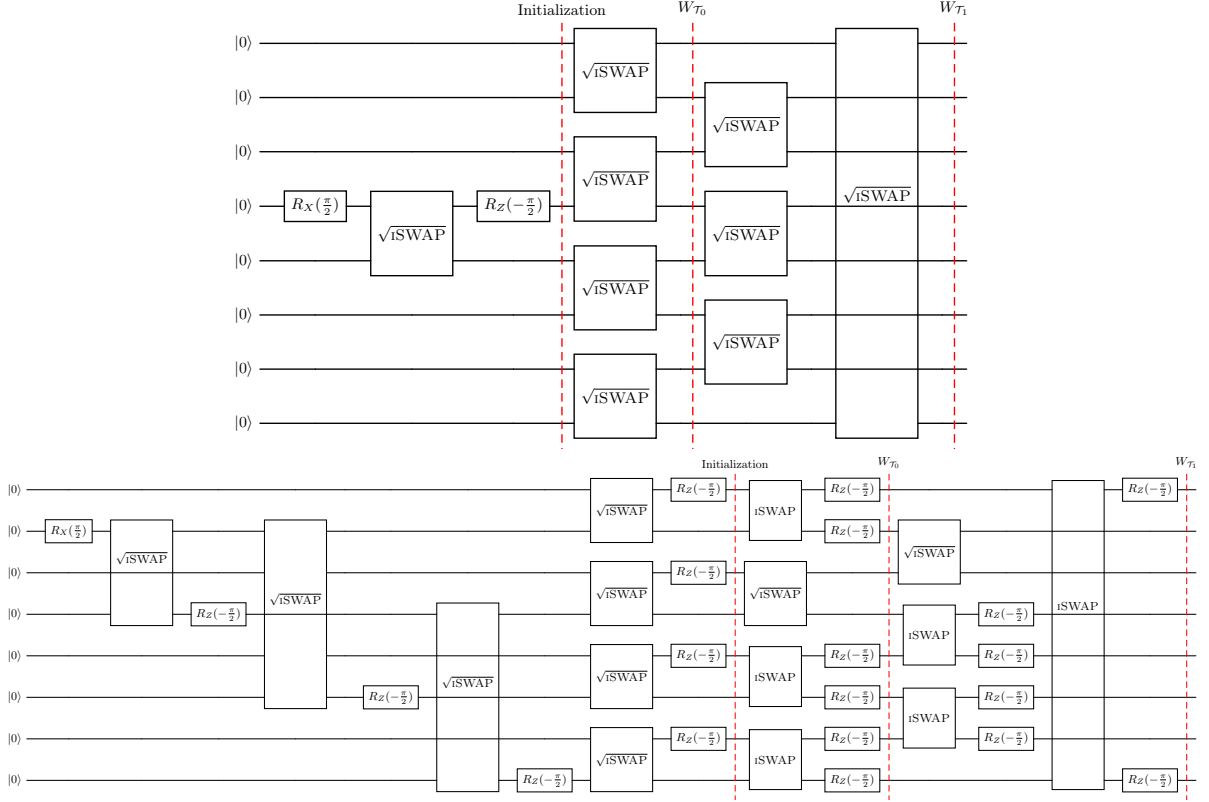

    \centering
    %\resizebox{10 cm}{!}{\input{1st_year/papers/im/latex_figs/QW_search}}
   % \caption{}
  %  \label{fig:circuits}
%\end{figure*}
%\begin{figure*}
 %   \centering
    \resizebox{10 cm}{!}{%
    \input{im/latex_figs/QW}}
    \resizebox{16 cm}{!}{%
    \input{im/latex_figs/QW_search_initial}}
    %\resizebox{10 cm}{!}{\input{1st_year/papers/im/latex_figs/QW_search}}
    \caption{Circuits implementing the initialization and one time-step of the QW (top) and QW search (bottom) via QCA over an 8-cycle. Time-steps are divided into $t_{\mathcal{T}_{0}}$ and $t_{\mathcal{T}_{1}}$, according to the tessellation cover defined in Eq. (\ref{eq:tess_1D}) and (\ref{eq:QCA_tr_func}), see Fig. \ref{fig:cycles_torus}. In the QW, the qubits are initialized in the state $\ket{\psi_{0}}=\frac{1}{\sqrt{2}}(\ket{3}+\ket{4})$, while in the QW search they are initialized in an equiprobable superposition of the allowed one-particle sector states (i.e. Dicke state with hamming weight one) $\frac{1}{\sqrt{2^{N}}}\sum_{i=0}^{N-1}\ket{\cdots,0_{i-1},1_{i},0_{i+1},\cdots}$. In the QW search, the marked vertex is mapped to the third qubit (third line of the circuit).}
    \label{fig:circuits}
\end{figure*}
%
%%%%%%%%%%%%%%%%%%%%%%%%%%%%%%%%%%%%%%%%%%%%%%%%%%%%%%%%%%%%%%%%%%
%bibliography
%%%%%%%%%%%%%%%%%%%%%%%%%%%%%%%%%%%%%%%%%%%%%%%%%%%%%%%%%%%%%%%%%%
%
\clearpage

%\bibliography{bib_quentin}
%apsrev4-2.bst 2019-01-14 (MD) hand-edited version of apsrev4-1.bst
%Control: key (0)
%Control: author (8) initials jnrlst
%Control: editor formatted (1) identically to author
%Control: production of article title (0) allowed
%Control: page (0) single
%Control: year (1) truncated
%Control: production of eprint (0) enabled
%

%

\end{document}